\documentclass[iop]{emulateapj}
\usepackage{amssymb,amsmath}
\newcounter{address}
\newcommand{\ie}{i.e.}

\newcommand{\degree}{\ensuremath{^\circ}}

\newcommand{\degsq}{\,deg$^{2}$}
\newcommand{\perdegsq}{\,deg$^{-2}$}
\newcommand\geqsim{\lower.73ex\hbox{$\sim$}\llap{\raise.4ex\hbox{$>$}}$\,$}
\newcommand\leqsim{\lower.73ex\hbox{$\sim$}\llap{\raise.4ex\hbox{$<$}}$\,$}
\newcommand{\project}[1]{\emph{#1}}
\newcommand{\sdss}{\project{SDSS}}

\newcommand{\dre}{\project{DR8}}

\newcommand{\sdssi}{\project{SDSS-I}}
\newcommand{\sdssii}{\project{SDSS-II}}
\newcommand{\sdssiii}{\project{SDSS-III}}
\newcommand{\sdssiv}{\project{SDSS-IV}}
\newcommand{\sequels}{\project{SEQUELS}}

\newcommand{\panstarrs}{\project{Pan-STARRS}}

\newcommand{\wise}{\project{WISE}}

\newcommand{\boss}{\project{BOSS}}
\newcommand{\eboss}{\project{eBOSS}}

\begin{document}

\title{The SDSS-IV extended Baryonic Oscillation Spectroscopic Survey: Luminous Red Galaxy Target Selection}
\author{
Abhishek~Prakash\altaffilmark{1},
Timothy~C.~Licquia\altaffilmark{1},
Jeffrey~A.~Newman\altaffilmark{1},
Ashley~J.~Ross\altaffilmark{2,6},
Adam~D.~Myers\altaffilmark{3},
Kyle~S.~Dawson\altaffilmark{4},
Jean-Paul~Kneib\altaffilmark{5,9},
Will~J.~Percival\altaffilmark{6},
Julian~E.~Bautista\altaffilmark{4},
Johan~Comparat\altaffilmark{7},
Jeremy~L.~Tinker\altaffilmark{8},
David~J.~Schlegel\altaffilmark{10},
Rita~Tojeiro\altaffilmark{11},
Shirley~Ho\altaffilmark{12},
Dustin~Lang\altaffilmark{12},
Sandhya~M.~Rao\altaffilmark{1},
Cameron~K.~McBride\altaffilmark{13},
Guangtun~Ben~Zhu\altaffilmark{14},
Joel~R.~Brownstein\altaffilmark{4},
Stephen~Bailey\altaffilmark{10},
Adam~S.~Bolton\altaffilmark{4},
Timoth\'ee~Delubac\altaffilmark{5},
Vivek~Mariappan\altaffilmark{4},
Michael~R.~Blanton\altaffilmark{8},
Beth~Reid\altaffilmark{15},
Donald~P.~Schneider\altaffilmark{16},
Hee-Jong~Seo\altaffilmark{17},
Aurelio~Carnero~Rosell\altaffilmark{18,19},
Francisco~Prada\altaffilmark{7,20,21}
}

\altaffiltext{1}{
PITT PACC, Department of Physics and Astronomy, 
University of Pittsburgh, Pittsburgh, PA 15260, USA.
}

\altaffiltext{2}{
Center for Cosmology and Astro-Particle Physics,
Ohio State University, Columbus, OH 43210
}

\altaffiltext{3}{
Department of Physics and Astronomy, 
University of Wyoming, 
Laramie, WY 82071, USA.
}

\altaffiltext{4}{
Department of Physics and Astronomy, 
University of Utah, Salt Lake City, UT 84112, USA.
}

\altaffiltext{5}{
Laboratoire d'astrophysique, Ecole Polytechnique F\'ed\'erale de Lausanne
Observatoire de Sauverny, 1290 Versoix, Switzerland
}

\altaffiltext{6}{
Institute of Cosmology \& Gravitation, Dennis Sciama Building, University of Portsmouth, Portsmouth, PO1 3FX, UK.
}

\altaffiltext{7}{
Instituto de F\'{\i}sica Te\'orica, (UAM/CSIC), 
Universidad Aut\'onoma de Madrid, Cantoblanco, E-28049 Madrid, Spain.
}

\altaffiltext{8}{
Center for Cosmology and Particle Physics,
Department of Physics, New York University,
4 Washington Place, New York, NY 10003, USA.
}

\altaffiltext{9}{
Aix Marseille Universit\'e, CNRS, LAM (Laboratoire d'Astrophysique de Marseille) UMR 7326, 13388, Marseille, France
}

\altaffiltext{10}{
Lawrence Berkeley National Laboratory, One Cyclotron Road,
Berkeley, CA 94720, USA.
}

\altaffiltext{11}{
School of Physics and Astronomy,
St Andrews
KY16 9SS
UK
}

\altaffiltext{12}{
Bruce and Astrid McWilliams Center for Cosmology,
Department of Physics, 
Carnegie Mellon University, 5000 Forbes Ave, Pittsburgh, PA 15213, USA.
}

\altaffiltext{13}{
Harvard-Smithsonian Center for Astrophysics,
Harvard University,
60 Garden St.,
Cambridge MA 02138, USA.
}

\altaffiltext{14}{
Department of Physics and Astronomy, Johns Hopkins University, Baltimore, MD 21218, USA.
}

\altaffiltext{15}{
Berkeley Center for Cosmological Physics,
LBL and Department of Physics,
University of California, Berkeley, CA 94720, USA.}

\altaffiltext{16}{
Department of Astronomy and Astrophysics, 525 Davey Laboratory, 
The Pennsylvania State University, University Park, PA 16802, USA.
}

\altaffiltext{17}{
Department of Physics and Astronomy,
Ohio University, Athens, OH 45701
}

\altaffiltext{18}{
Observat\'orio Nacional, Rua Gal. Jos\'e Cristino 77, 
Rio de Janeiro, RJ - 20921-400, Brazil.
}

\altaffiltext{19}{
Laborat\'orio Interinstitucional de e-Astronomia, - LIneA, 
Rua Gal.~Jos\'e Cristino 77, 
Rio de Janeiro, RJ - 20921-400, Brazil.  
}

\altaffiltext{20}{
Campus of International Excellence UAM+CSIC, Cantoblanco, E-28049 Madrid, Spain
}

\altaffiltext{21}{
Instituto de Astrof\'{\i}sica de Andaluc\'{\i}a (CSIC), Glorieta de la Astronom\'{\i}a, E-18080 Granada, Spain
}
\email{abp15@pitt.edu}

\begin{abstract}
We describe the algorithm used to select the Luminous Red Galaxy (LRG) sample for the {\em extended Baryon Oscillation Spectroscopic Survey} (\eboss) of the Sloan Digital Sky Survey IV (SDSS-IV) using photometric data from both the SDSS and the Wide-Field Infrared Survey Explorer (WISE).  LRG targets are required to meet a set of color selection criteria and have \textit{z}-band and \textit{i}-band \texttt{MODEL} magnitudes $z < 19.95$ and  $19.9 < i < 21.8$, respectively. Our algorithm selects roughly 50 LRG targets per square degree, the great majority of which lie in the redshift range $0.6 < z < 1.0$ (median redshift 0.71). We demonstrate that our methods are highly effective at eliminating stellar contamination and lower-redshift galaxies. We perform a number of tests using spectroscopic data from SDSS-III/\boss\ to determine the redshift reliability of our target selection and its ability to meet the science requirements of \eboss. The SDSS spectra are of high enough signal-to-noise ratio that at least $\sim89\%$ of the target sample yields secure redshift measurements. We also present tests of the uniformity and homogeneity of the sample, demonstrating that it should be clean enough for studies of the large-scale structure of the universe at higher redshifts than  SDSS-III/\boss\ LRGs reached.  
\end{abstract}


\section{Introduction}\label{sec:intro}

Many studies have found that massive galaxies, and in particular red, elliptical galaxies, tend to reside in massive dark matter halos and cluster strongly \citep[e.g.,][]{postman_geller1984,Kauffmann2004-353}. The most luminous galaxies in clusters and groups populate a narrow range of color and intrinsic luminosity \citep{postman_and_lauer1995}. These galaxies, which constitute the most massive, the most luminous, and the reddest (in rest-frame color) of all galaxies are typically referred to as `Luminous Red Galaxies' or LRGs. Given both their bright intrinsic luminosities (allowing them to be studied to higher redshifts than typical $L^{*}$ galaxies) and their strong clustering, LRGs are excellent tracers of the large-scale structure of the Universe.

LRGs have been previously used to study large-scale structure by a variety of investigations, most notably the Sloan Digital Sky Survey (SDSS; \citet{york_sdss_2000}) and the \sdssiii /Baryon Oscillation Spectroscopic Survey (\boss), as well as the 2dF-SDSS LRG and QSO survey \citep[e.g.][]{Eisenstein2001,Cannon2006}.  In combination, \sdssi, \sdssii\ and \sdssiii\ targeted LRGs at $z~\leqsim\,0.7$ to a magnitude limit of $i < 19.9$ and $i_{fiber2} < 21.5$  \citep{Eisenstein2001,Eisenstein2005,eisenstein_sdss3_2011,dawson_boss_2013}.  The methods used to select LRGs for these studies are limited in redshift range as a result of using optical photometry alone for selection.  Identifying LRGs with shallow optical photometry becomes prohibitively difficult at higher redshifts as the 4000\,\AA\  break passes into the near-infrared and colors  overlap strongly with M stars.

New multi-wavelength imaging is now available which allows high-redshift LRGs to be selected much more efficiently than optical-only imaging would make possible. In particular, optical-infrared (optical-IR) colors provide a powerful diagnostic for separating galaxies and stars\citep{abhi_2015a}, as well as a diagnostic of redshift.  As a result, infrared observations from satellites such as the Wide-field Infrared Survey Explorer \citep[\wise;][]{wright2010} provides additional information for targeting LRGs in regions of optical color space that would otherwise be heavily contaminated by stars.

Increasing our current sample of LRGs to higher redshifts will allow measurements of the Baryon Acoustic Oscillation (BAO) feature, and hence of the expansion rate of the Universe \citep{Seo03,Lin03,nic_lrg_2008}, during the era when accelerated expansion began. An optical + \wise\ selection makes it possible to target LRGs in the redshift range  $0.6~\leqsim\,z~\leqsim\,1$ efficiently \citep{abhi_2015a}; with spectroscopy of these targets, we can obtain stronger constraints on the BAO scale at these redshifts. At even higher redshift, other tracers such as quasi-stellar objects (quasars) and Emission Line Galaxies (ELGs) can be used to provide further complementary probes of the BAO scale. In combination, these target classes can provide powerful constraints on the evolution of cosmic acceleration across a wide range of redshifts. This led to the conception of a new survey, the {\em extended Baryon Oscillation Spectroscopic Survey} (\eboss; Dawson et al. 2015) as part of \sdssiv (Blanton et al. in prep.).

The LRG component of \eboss\ will obtain spectra for $\sim$375{,}000 objects. Approximately 265{,}000 of these are expected to be LRGs in the redshift range $0.6 < z < 1.0$, with a median redshift of $z\sim0.7$. The main goal of this spectroscopic campaign is to produce more precise measurements of the BAO signal at $0.6 < z < 1.0$, thus extending probes of the BAO scale using LRGs beyond the \boss\ redshift range. \eboss\ LRGs are also expected to yield a $4\%$ measurement of Redshift Space Distortions (RSD) which will allow improved tests of General Relativity at these redshifts\citep[e.g.,][]{Beutler14a,Beutler14b,Samushia14}.

Altogether, \sdssiv/\eboss\ will produce a spectroscopic sample of both galaxies and quasars over a volume that is 10 times larger than the final \sdssiii\ \boss\ sample, although at lower target density. This sample will enable a wide range of scientific studies beyond a BAO measurement. For example, the resulting sample of hundreds of thousands of LRGs extending to $z = 1$ will be ideal for evolution studies of the brightest elliptical galaxies, including measurements of luminosity functions, mass functions, size evolution, and galaxy lensing.

In this paper, we describe the algorithm used to select LRG targets for the \eboss\ survey.
Further technical details about \eboss\ can be found in companion papers on  
Quasar selection \citep{eboss_qso}, ELG selection \citep{Comparat_elg}, survey strategy \citep{eboss_overview}, and the {\it Tractor} analysis of \wise\ data \citep{Lan14}.

The paper is organized as follows. In \S2 we outline the goals of \eboss\ and the requirements placed on the LRG sample to meet these goals. The parent imaging data used for \eboss\ LRG target selection 
is outlined in \S3. In \S4 we describe our new method of LRG selection and supporting tests for this method that were conducted during \boss. In \S5 we describe the \eboss\ LRG targeting algorithms and the meaning
of the relevant targeting bits, while \S6 uses the latest results from \eboss\ to test the target selection algorithm. An important criterion for any large-scale structure survey is sufficient homogeneity to facilitate modeling of the distribution of the tracer 
population, i.e., the `mask' of the survey. In \S7, we use the full \eboss\ target sample 
to characterize the homogeneity of \eboss\ LRGs. We present conclusions and future implications for \eboss\ LRGs in \S8. 

Unless stated otherwise, all magnitudes and fluxes in this paper are corrected for extinction
using the dust maps of \citet{Sfd1998}, hereafter SFD, and are expressed in the AB system \citep{oke_gunn_1983}. The SDSS photometry has been demonstrated to have colors that are within 3\% of being on an AB system \citet{snf_sdss2011}. We use a standard $\Lambda$CDM cosmology with $H_{0}$=100\textit{h} km s$^{-1}$ Mpc$^{-1}$, $h=0.7$, $\Omega_{\rm M} = 0.3$, and $\Omega_\Lambda = 0.7$, which is broadly consistent with the recent results from {\em Planck} \citep[][]{Planck14}.

\section{Cosmological Goals of eBOSS and Implications for LRG Target Selection}\label{sec:eboss_goals}
\subsection{Overall Goals for the Luminous Red Galaxy Sample}\label{sec:lrg_eboss_goals}
The primary scientific goals of the \eboss\ LRG survey are to constrain  the scale of the BAO to $1\%$ accuracy over the redshift regime $0.6 < z < 1.0$. This requires selecting a statistically uniform set of galaxies with the desired physical properties for which spectroscopic redshifts can be efficiently measured. The density of selected LRGs must not strongly correlate with either tracers of potential imaging systematics (e.g., variations in the depth of the imaging) or with astrophysical systematics such as Galactic extinction and stellar density.

\subsection{Target Requirements for LRGs}\label{sec:ts_req}
As explained in \citet{eboss_overview}, a density of 50 deg$^{-2}$ spectroscopic fibers are allocated to \eboss\ LRGs. Thus, any resulting sample will have approximately 1/3 the number density of the $z < 0.6$ BOSS sample. We therefore expect \eboss\ LRGs to have a bias of $1.7\sigma_8(0)/\sigma_8(z)$, assuming the sample contains objects that are, statistically speaking, the progenitors of the brightest 1/3 of \boss\ LRGs (see, e.g., \citealt{Ross14}). Under this assumption, a density of 40 deg$^{-2}$ LRGs with redshifts $0.6 < z < 1.0$ is required, over the projected 7000 deg$^{2}$ survey footprint, to meet the \eboss\ scientific goals (see \citet{eboss_overview} for more details). This corresponds to a requirement that 80\% of \eboss~ LRG targets result in a spectroscopically confirmed galaxy with $0.6 < z < 1.0$, with a median redshift $z>0.7$. Additionally, we require the redshifts be accurate to better than 300 km s$^{-1}$ RMS and robust such that the fraction of catastrophic redshift errors (exceeding 1000 km s$^{-1}$) is $< 1$\% in cases where the redshifts are believed to be secure. The construction of a sample designed to fulfill these requirements is described in Sections \ref{sec:cat} and \ref{sec:target_alg}.\\
\indent A further requirement to obtain robust BAO and RSD measurements is that the density of selected LRGs must not strongly correlate with either tracers of potential imaging systematics (e.g., variations in the depth of the imaging) or with astrophysical systematics such as Galactic extinction and stellar density.\boss\ has shown that fluctuations associated with surveys artifacts can be handled effectively via weighting schemes provided the amplitude of fluctuations is relatively small \citep{ashley2012}. To facilitate weighting schemes in future clustering studies, we require that that fluctuations in the expected target density as a function of potential imaging systematics, stellar density, and Galactic extinction be less than 15\%\ (total variation around mean density).  We require density differences due to imaging zero point variations in any single band to be below 15\% as well. Tests of the homogeneity of the LRG target sample are presented in Section \ref{sec:uni_test}.

\section{Parent Imaging for Target Selection}\label{sec:imaging}
\subsection{Updated calibrations of \sdss\ imaging}
\label{sec:v5b}

All \eboss\ LRG targets rely on imaging from the \sdss-\project{I}/\project{II}/\project{III}.
 \sdss\ photometry was obtained by the \sdss\ telescope \citep{gunn_sdss_2006} using its 
wide-field imaging camera \citep{gunn_sdss_1998} in the $ugriz$ system \citep{fukugita_sdss_1996}. 
\sdss-\project{I}/\project{II} primarily obtained imaging over the $\sim8400$\degsq\ ``Legacy'' area, 
$\sim90$\% of which was in the North Galactic Cap (NGC). This imaging was
released as part of \sdss\ Data Release 7 (DR7;\citet{DR7}). The legacy imaging
area of the \sdss\ was expanded by $\sim2500$\degsq\ in the South Galactic Cap (SGC) as 
part of \dre\ \citep{DR8}. The \sdss-\project{III}/\boss\ survey used this DR8 imaging for target
selection over $\sim7600$\degsq\  in the NGC and $\sim3200$\degsq\  in the 
SGC \citep{dawson_boss_2013}. LRG targets for \eboss\ have been selected over the same footprint covered by \boss;
ultimately \eboss\ will obtain spectroscopy for LRGs over a roughly 7500\degsq\ subset of this \boss\ area.

Although conducted over the same {\em area} as \boss, \eboss\ target selection
takes advantage of updated calibrations of the \sdss\ imaging. 
\citet{Sch12} have applied the ``uber-calibration'' 
technique of \citet{pad_sdss_2008} to imaging from the \panstarrs\ survey \citep{kaiser_2010},
achieving an improved global calibration compared to \sdss\ \dre. Targeting
for \eboss\ is conducted using \sdss\ imaging that is calibrated using the \cite{Sch12} \panstarrs\ solution.
We will refer to this as the ``updated'' photometry below.
 
Specifically, targets are selected using the updated \sdss\ photometry stored in the {\tt calib\_obj} files, the basic imaging catalog files used in the \sdss-\project{III} data model.\footnote{e.g., \url{http://data.sdss3.org/datamodel/files/PHOTO\_SWEEP\\/RERUN/calibObj.html}}  
The updated Pan-STARRS-calibrated photometry will be made available as part of a future \sdss\ Data Release. 
The magnitudes provided in these files are Pogson magnitudes \citep{pogson_mags_1968} rather than the asinh magnitudes used for some \sdss\ data releases \citep{lup_sdss_1999}. We use \texttt{Model Magnitudes} for all colors and fluxes used in selection. The \texttt{Model Magnitudes} are obtained by first determining what type of model (exponential or deVaucouleurs) best fits the object image in the 'canonical' band (typically \textit{r}, but other bands may be used if they have higher signal-to-noise), and then using the model fit from the canonical band (convolved with the appropriate PSF) to obtain fluxes in each filter. Additionally, we also apply flux limits based upon an object's \texttt{fiber2mag} values; i.e., the total flux within a 2'' diameter of the object center, corresponding to the aperture of a \boss\ spectroscopic fiber \citep{smee_sdss_2013}.

\subsection{WISE}
 The \eboss\ LRG target selection algorithm also relies on infrared photometry from the Wide-Field Infrared Survey Explorer \citep[\wise;][]{wright2010}. \wise\ observed the full sky in four infrared channels centered at 3.4, 4.6, 12, and 22 microns, which we refer to as W1, W2, W3, and W4, respectively. For \eboss\ LRGs, we use the W1 band only. \wise\ magnitudes are commonly measured in the Vega system, but we convert to the AB system for LRG selection.\footnote{$W1_{\rm AB} = W1_{\rm Vega} + 2.699$} Over the course of its primary mission and the `NEOWISE post-cryo' continuation, \wise\ completed two full scans of the sky
in the W1 and W2 bands. Over 99\% of the sky has 23 or more exposures in W1 and W2, and the median coverage is 33 exposures. We use the `unWISE' forced photometry from \citet{Lan14}, which photometered custom coadds of the \wise\ imaging at the positions of all \sdss\ primary sources. Using forced photometry allows accurate flux measurements to be obtained even for significantly blended sources, including objects below the significance threshold for \wise\-only detections. Since the \wise\ W1 point-spread function is relatively broad (6.1 arc-seconds FWHM, $\sim$4 times larger than typical \sdss\ seeing), many sources are blended and forced photometry presents substantial advantages. Additionally, forced photometry allows us to leverage the relatively deep \sdss\ photometry to measure fluxes of \wise\ sources that are otherwise below the detection threshold.  Using unWISE photometry instead of the \citet{wright2010} \wise\ catalog increases the size of the resulting \eboss\ LRG sample by $\sim10\%$. 

\section{Selection of High-$z$ LRGs}\label{sec:cat}

 Our overall goal is to cleanly select a sample of LRGs at redshifts beyond 0.6.  In this redshift regime, however, optical photometry alone becomes insufficient for discriminating these high-$z$ objects from foreground stars in our Galaxy because both LRGs and red stars occupy the same region in optical color-color space.  \citet{abhi_2015a} presented a new technique which eliminates almost all stellar contamination by combining both optical and infrared imaging data. This takes advantage of the prominent 1.6 micron `bump' in the spectral energy distributions of LRGs and other objects with old stellar populations \citep{john1988}, which results from the minimum in the opacity of H$^{-}$ ions. The lowest wavelength channel of the \wise\ satellite is centered at 3.4 microns, almost perfectly in sync with the bump at $z\sim1$.
 
\begin{figure}[btp]
\centering
	\includegraphics[scale=0.20]{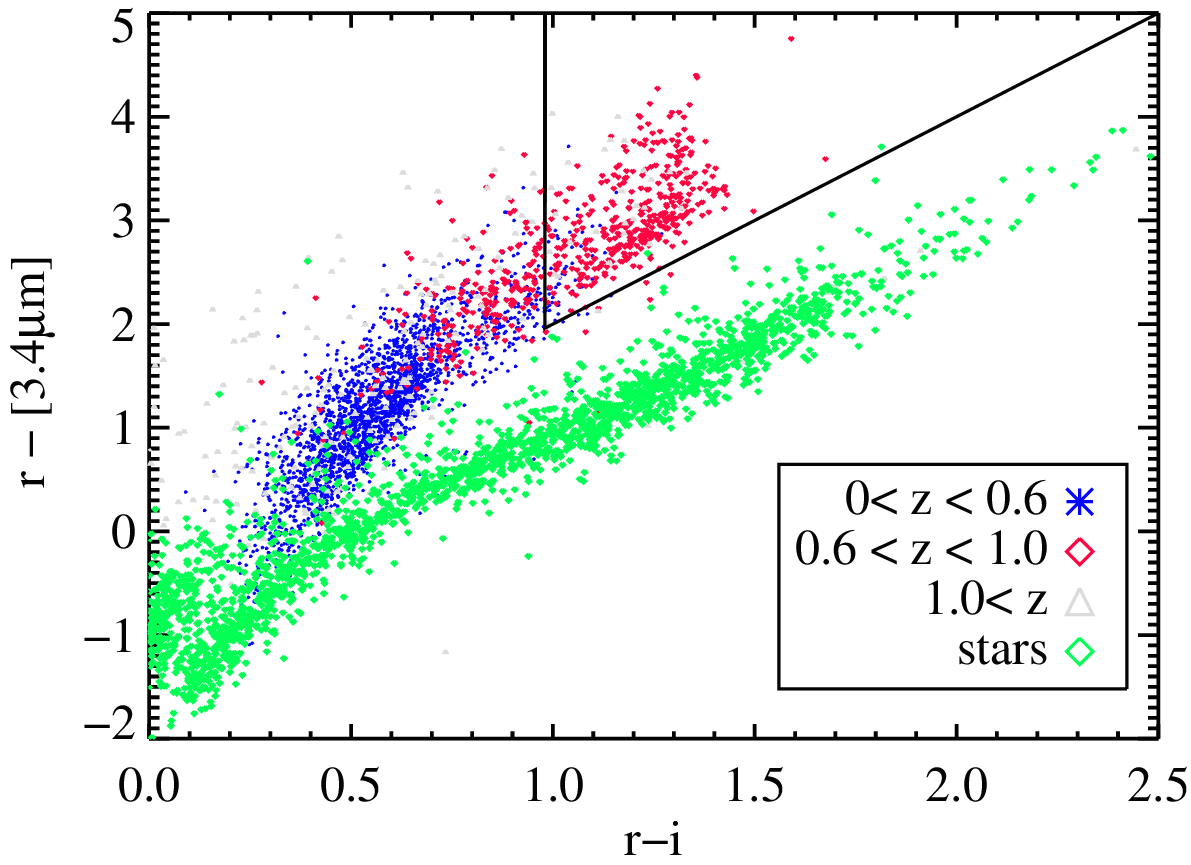}
	\caption{  {Optical/infrared color-color plot for galaxies observed by WISE and CFHT\,LS with photometric redshifts from the COSMOS survey. Blue symbols represent galaxies with photometric redshifts of z $<$ 0.6, red diamonds represent galaxies at  0.6 $<$ z $<$ 1.0, and cyan triangles represent galaxies at z $>$ 1.0. Stars are represented by green diamonds. The triangular area depicts the broad selection presented in Equations~\ref{eqn:nom_cuts1} and~\ref{eqn:nom_cuts2}. Photometric redshifts are taken from the COSMOS photo-z catalog of \citet{ilbert_2008} and optical photometry is from the catalog of \citet{gwyn_2011}, transformed to SDSS passbands. The conversion relation can be found at CFHT LS webpage.\footnote{http://www.cadc-ccda.hia-iha.nrc-cnrc.gc.ca/en/megapipe/docs/filt.html} }}
\label{fig:rw1_cos}
\end{figure}

Figure~\ref{fig:rw1_cos} shows both stars and galaxies in a plot of \textit{r-W1} verses \textit{r-i} color, where \textit{W1} indicates the magnitude of a source in the \wise\ 3.4 micron pass-band (on the AB system) and \textit{r} and \textit{i} indicate \sdss\ model magnitudes in the appropriate passband.  Stars separate increasingly from the galaxy population in near-IR/optical color space as redshift increases, allowing clean discrimination of galaxies at \textit{z} $>$ 0.6 from stars. Simultaneously, \textit{r-i} color increases with increasing redshift (particularly for intrinsically red galaxies) as the 4000\,\AA\ break shifts redward, allowing a selection specifically for higher-redshift objects. While the combination of optical and IR imaging provides an excellent means of removing stellar contamination from an LRG target sample, this approach also means that we are limited to objects that are detected by {\em both} \sdss\ and \wise. Using forced photometry enables a more complete LRG sample by allowing objects which are poorly detected in one dataset or the other to still be selected.

As a basic color selection for characterizing potential \eboss\ LRG targets, we select all objects that satisfy the criteria
\begin{align}
      \textit{r-i} > 0.98, {\rm and}\label{eqn:nom_cuts1}\\
       \textit{r-W1} > 2.0 \times (r-i), \label{eqn:nom_cuts2}
\end{align}
where all magnitudes are corrected for Galactic extinction. These cuts were determined by examining the location of objects of known redshift and restframe color in color-color space, as in Figure~\ref{fig:rw1_cos}. Further details on the motivation for this selection can be found in \cite{abhi_2015a}.

\begin{figure*}[btp]
\centering
	\includegraphics[scale=0.62]{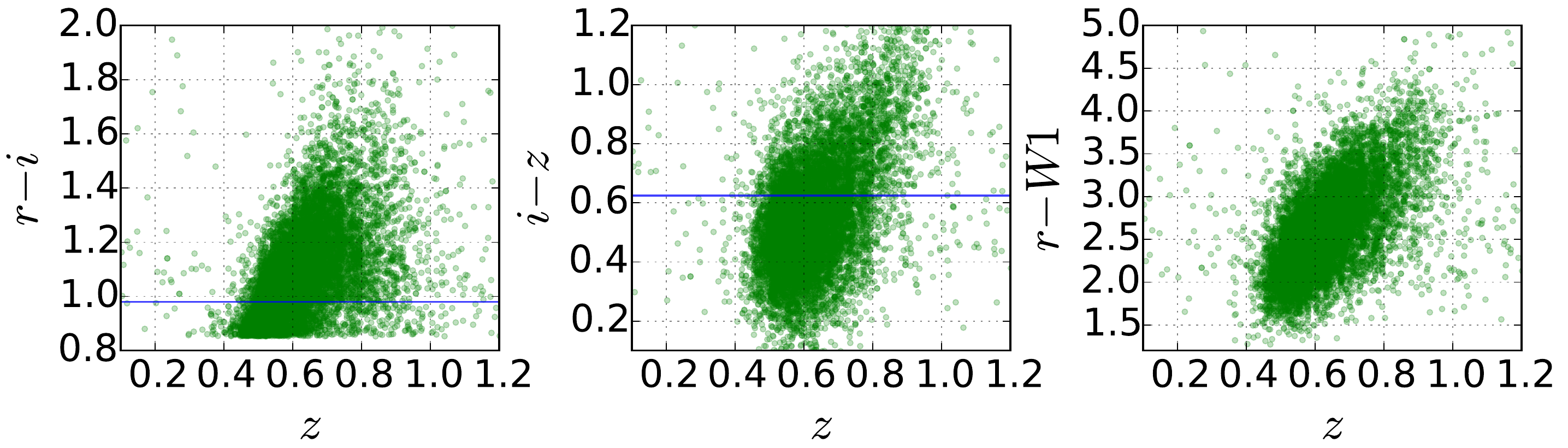}
	\caption{Plots of $r-i$, $i-z$ and $r-W1$ color as a function of redshift for 15,000 LRGs targeted via a \boss\ ancillary 
program \citep[see the Appendix of][]{DR12}, which utilized broader selection criteria than those used for \eboss. The blue lines represent the cuts applied as part of the \eboss\ target selection algorithm. Selecting objects with r-i $>$ 0.98, i-z $>$ 0.625 and $r-w1 >2*(r-i)$ rejects a significant number of $z < 0.6$ galaxies while missing relatively few $z > 0.6$ LRGs.
}
\label{fig:lrg_zhist_wake}
\end{figure*}

To test this new selection technique, we targeted 10{,}000 objects satisfying this selection in a \boss\ ancillary program in 2012-2013 \citep[see the Appendix of][]{DR12}. Selection was limited to objects with $z_{Model}  < 20$; 98\% of the spectra yielded secure redshift measurements. These redshift estimates were found to be reproducible when observed multiple times. An additional 5{,}000 LRGs were selected by relaxing the \textit{r-i} color requirement to \textit{r-i} $>$ 0.85 in order to estimate the number of LRGs missed by the color cuts in Equation~\ref{eqn:nom_cuts1}. The distributions of observed colors as a function of redshift for the resulting sample of 15,000 LRGs is presented in Figure~\ref{fig:lrg_zhist_wake}.

Our method of combining optical and infrared photometry for this selection is unique; however, the specific choice of color cuts is not. We are able to cleanly select similar samples of LRGs by using different color combinations; e.g., \textit{r-W1} and \textit{r-z}, or \textit{i-W1} and \textit{i-z}. As can be seen in Figure~\ref{fig:lrg_zhist_wake}, incorporating multiple colors can improve the efficiency of identifying true LRGs in the redshift range of interest by rejecting lower-redshift objects.

\section{The \eboss\ LRG target selection algorithm}\label{sec:target_alg}
In this section, we describe in detail the final selection algorithm for \sdssiv/\eboss\ LRGs. At the high redshifts of the LRGs ($z$ $>$ 0.6), the 4000\,\AA\ break moves into the \sdss\ \textit{i}-band. This causes the flux measured in the \textit{r}-band to be negative occasionally, so the color cuts are made in flux space rather than magnitude space. However, for convenience, we describe the selection algorithm and flux limits in terms of extinction-corrected AB magnitudes and colors here. 

To summarize our selection methods: We first employ photometric processing flags to eliminate those objects with problematic imaging.\footnote{\url{https://www.sdss3.org/dr8/algorithms/photo\_flags\_recommend.php}}  To ensure robust selection while maintaining a sufficient signal-to-noise ratio in \eboss\ spectra, we also apply a variety of flux limits.  Finally, to maximize the fraction of targets that are in fact high-redshift LRGs, we apply several color cuts.  In the following sub-sections, we detail all the selections used for creating the \eboss\ LRG sample.

\subsection{Photometric flags for the LRG sample}\label{sec:phot_flag_eboss}

Since many of the \sdss\ imaging runs overlap on the sky, an object may be observed twice or more \citep{stoughton_sdss_2002}. Only one observation is designated as the \texttt{primary} observation of the object during the \texttt{resolve} process. Hence, to exclude duplicate objects we enforce the following logical condition on the {\tt RESOLVE\_STATUS} bit-mask:

\begin{eqnarray}
      (Resolve\_status\ \&\ Survey\_primary) \neq 0.\label{eqn:resolve_primary}
\end{eqnarray}
 No other SDSS imaging flags are used for selecting this sample, as most become uninformative for distinguishing between real  and spurious detections at the faint limits of our selection.  As a result, some false sources will make it into the \eboss\ LRG sample; however, once spectroscopy is obtained they may be eliminated from the sample.
\subsection{Magnitude limits}\label{sec:mag_lim}
The median 5-$\sigma$ depth for photometric observations of point sources in the \sdss\ is $u = 22.15$, $g = 23.13$, $r = 22.70$, $i = 22.20$, $z = 20.71$ \citep{eboss_overview}. We further require a detection of the flux in the \textit{W1} forced photometry for an object to be targeted. Keeping these requirements in mind, we apply the following flux limits to the entire sample:

\begin{eqnarray}
      MODEL\_IVAR_{r,i,z} \neq 0,\label{eqn:flux_err}\\     
      z_{Fiber2} \leq 21.7,\label{eqn:fiber2_lim}\\
      19.9 \leq i_{Model }  \leq 21.8,\label{eqn:i_mod_lim}\\
      z_{Model} \leq 19.95,\label{eqn:z_mod_bright_lrg}\\   
      W1_{vega} \neq 0, {\rm and} \label{eqn:wise_mag}\\
      W1_{AB}  \leq 20.299, \label{eqn:wise_flux_lim}
\end{eqnarray}

where \texttt{MODEL\_IVAR} are the inverse variances on the model fluxes in \textit{r}, \textit{i}, and \textit{z} bands. The application of Equation~\ref{eqn:fiber2_lim} serves to maintain a sufficiently high signal-to-noise ratio of the \eboss\ spectra. This cut is similar in spirit to the $i_{Fiber2}$ cut that was used for the \boss\ CMASS galaxy sample \citep{eisenstein_sdss3_2011}. We apply the lower limit defined in Equation~\ref{eqn:i_mod_lim} in order to avoid targeting $i < 19.9$ \boss\ CMASS galaxies, which generally lie at lower redshifts and have been observed previously. $W1_{vega}$ being nonzero implies that the photometry is reliable, while Equation~\ref{eqn:wise_flux_lim} ensures that \wise\ flux measurements have a signal-to-noise ratio greater than 5\citep{wright2010}.   The \textit{i} and \textit{z} faint magnitude limits are set to achieve the required target density of $\sim$60 targets \perdegsq\, matching the \eboss\ fiber allocation for LRGs \citep[Dawson et al. 2015][]{}, while maximizing the brightness of targets.

\subsection{Color Selection}\label{sec:color_cut}
 We use the \textit{r-W1} (optical-infrared) color for separating LRGs from stars.\footnote{Note that we do not explicitly use any morphological cuts, but rather separate stars and galaxies based only on their colors.} The optical colors of galaxies are used to ensure that the targeted objects are intrinsically red and lie in the desired redshift range. We thus apply the following three selection criteria:
\begin{eqnarray}   
      r-i > 0.98, \label{eqn:ri_bright_lrg}\\
      r-W1 > 2.0 \times (r-i), {\rm and} \label{eqn:rw1_bright_lrg}\\
      i-z > 0.625. \label{eqn:iz_bright_lrg}
\end{eqnarray}
Equations~\ref{eqn:ri_bright_lrg} and ~\ref{eqn:rw1_bright_lrg} represent the basic LRG color selection discussed at the beginning of \S\ref{sec:cat} and are identical to Equations~\ref{eqn:nom_cuts1} and ~\ref{eqn:nom_cuts2}, the color curs used in initial tests of LRG selection. We use Equation~\ref{eqn:iz_bright_lrg} to reduce contamination from $z < 0.6$ galaxies. 

\begin{figure*}[btp]
\centering
	\includegraphics[scale=0.77]{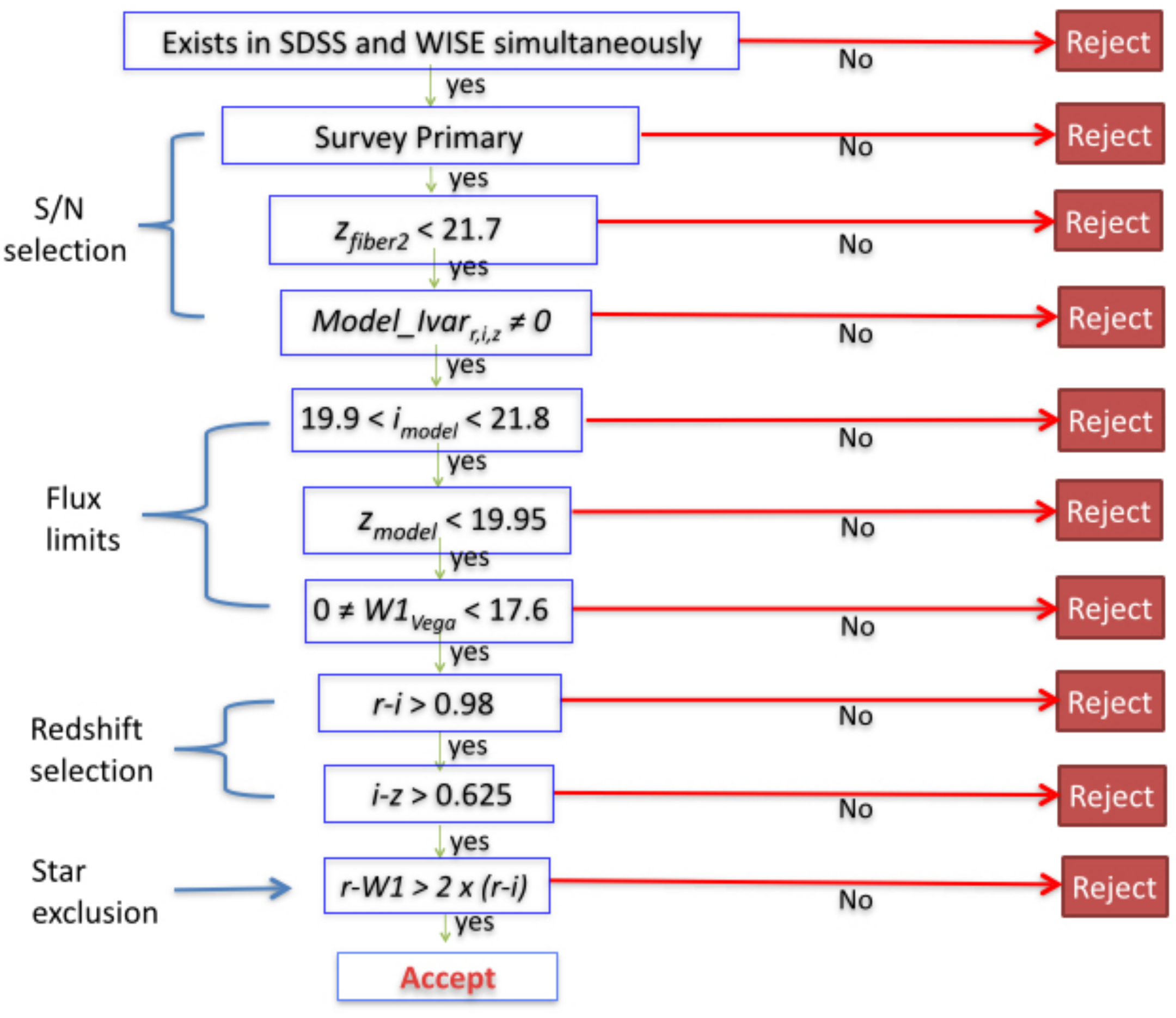}
	\caption{Schematic flow chart for the \eboss\ LRG target selection algorithm. All quantities are corrected for Galactic extinction. Refer to the text for a full description of all of the quantities shown in this figure.}
\label{fig:lrg_fc}
\end{figure*}

The overall \eboss\ LRG selection algorithm is shown schematically as a flow chart in Figure~\ref{fig:lrg_fc}. The details of this algorithm were optimized based upon a pilot survey, the {\em Sloan Extended Quasar, ELG and LRG Survey} (\sequels), which is summarized in the appendix of Alam et al. (2015); the \sequels\ LRG selection algorithm is detailed in an Appendix (see \S\ref{sec:sequels}). 

In addition to the LRGs targeted by Equations~\ref{eqn:resolve_primary}--\ref{eqn:iz_bright_lrg}, we target a small number of objects, $\sim$200 over the 10,000 \degsq\ SDSS imaging area, via a different but related algorithm. These objects have ${i_{Model}}$ $\geq$ 21.8 and are designated \texttt{LRG\_IDROP}. These are not significant for BAO studies but constitute a separate sample designed to identify rare objects at extremely high redshifts. 
Further details are provided in an Appendix to this paper (see \S\ref{sec:appendix_a}).

\section{Tests of the target selection algorithm}\label{sec:tsa_test}

In this section, we assess the results of our target selection methods using the current \eboss\ data. We use the automated spectral classification, redshift determination, and parameter measurement
pipelines of \sdssiii\ \boss\, which are described in \citet{bolton2012}, to reduce and analyze spectra of \eboss\ targets. We adopt the redshifts output as \texttt{z\_NOQSO}
by this pipeline, which corresponds to the set of chi-squared minima which are based on the assumption that an object is {\em not} a quasar, and hence must match stellar or galactic templates. To assess the true redshifts of LRG sample, we have conducted a visual inspection of a subset of \eboss\ spectra, employing the \texttt{idlspec2d} package for this purpose.\footnote{http://www.sdss3.org/dr8/software/products.php}

Specifically, we present results based on 2,557 LRG candidates from eight plates that were visually inspected to assess the quality of spectra and robustness of redshift measurements by a team of \eboss\ members. Each plate was inspected by multiple individuals to cross-check the results. Visual inspectors selected what they believed to be the best estimate of the correct redshift for each spectrum, as well as assessing the security of that redshift according to a simple four level confidence metric, \texttt{z\_conf} (confidence of inspector in the measured redshift). Targets are assigned \texttt{z\_conf} values of 0 to 3, with 2 and 3 corresponding to measurements which were believed to be robust. A value $z_{\rm conf} = 1$ denotes a spectrum that is ambiguously classified, i.e., where more than one of the chi-squared minima correspond to models which are a possible fit, while $\texttt{z\_conf} = 0 $ is used for objects where it is not possible to classify the objects and establish their redshift. These objects are considered unreliable and not used in the calculations of redshift distributions or related quantities.  

In the remainder of this section, we briefly present the expected basic characteristics of the \eboss\ LRG sample (e.g., its redshift distribution, spectral quality, and redshift success) derived from this sample with visual inspections. We also test the efficiency of our target selection algorithm against the science requirements for the \eboss\ LRG sample as described in \S\ref{sec:eboss_goals}. 

Two redshift distributions are presented in Table~\ref{table:sequels_eboss_riz_req}. The more conservative estimate (the one with a higher rate of ``Poor spectra'') assumes that only objects given $\texttt{z\_conf} > 1$ have been assigned a correct redshift.  The less conservative estimate includes all objects with $\texttt{z\_conf} >0$; this is a relevant scenario, since it is likely that a great majority of $\texttt{z\_conf}=1$ redshifts are correct, but will inevitably include at least some incorrect redshifts.  It is likely that the true distribution lies between these two bounds.  It is expected that pipeline improvements now underway will enable at least some redshifts currently assigned $\texttt{z\_conf}<2$ to be recovered automatically in the future.

As can be seen in the Table~\ref{table:sequels_eboss_riz_req}, even with the conservative scenario ($\texttt{z\_conf} >1$), the \sdss\ spectral pipeline generates a secure redshift solution for $\sim89\%$ of the LRG candidates visually inspected.  However, the fit determined to be correct via visual inspection sometimes does not correspond to the minimum chi-squared solution from the pipeline, but rather an alternative chi-squared minimum.\footnote{The \sdss\ pipeline generates a set of possible fits; cf. \citet{bolton2012}.}   Pipeline improvements now under way (which include both improved two-dimensional extractions and reductions in the freedom of template+polynomial fitting) are expected to improve the automated redshift-finding, so this figure should be a floor to the actual performance of \eboss\ LRGs. 

The remaining $\sim11-12\%$ of the LRG targets without a secure redshift determination typically have spectra with low signal-to-noise ratios.  An additional $\sim 9\%$ of the LRG targets are found to be stars. These two factors alone make it impossible to meet the LRG requirement that $80\%$ of all targets be LRGs within the range $0.6~\leqsim\,z~\leqsim\,1.0$, even before the redshift distribution of the galaxies is considered. In the end, 68--72\% of all LRG targets are in fact galaxies with definitive redshift measurements that lie in the desired regime. For detailed discussion of the pipeline results, visual inspections, templates and sources of redshift failures, see \citet{eboss_overview}.

 \begin{deluxetable}{lcr}
\tablecaption{Redshift distribution of \eboss\ LRGs, based upon results for a sample of 2,557 visually inspected spectra.
The surface densities are presented in units of deg$^{-2}$, normalizing to the total surface density of the parent sample for these spectra. Entries highlighted in bold font denote the subset of the sample that lies in the redshift range used to assess the high-level
science requirements for the LRG sample.}
\tablehead{ & \textbf{ LRGs}   & \textbf{ LRGs}\cr \\
 & $\texttt{z\_conf} >0$ & \ \ \ $\texttt{z\_conf} >1$ }
\startdata
Poor spectra ($\texttt{z\_conf} =0$) &  4.0 & 6.7   \cr
Stellar & 5.3  & 5.3   \cr
Galaxy &  N/A  &   N/A \cr
$0.0 < z < 0.5$ & 0.6 & 0.6    \cr
$0.5 < z < 0.6$ & 6.2 & 5.9     \cr
$0.6 < z < 0.7$ & {\bf 15.2} & {\bf 14.8}   \cr
$0.7 < z < 0.8$ & {\bf 15.3} & {\bf 14.7}   \cr
$0.8 < z < 0.9$ & {\bf 9.4} &   {\bf 8.7}      \cr
$0.9 < z < 1.0$ & {\bf 3.2} & {\bf 2.7}   \cr
$1.0 < z < 1.2$ & 0.6 & 0.5   \cr \hline
Targets  & 60 & 60 \cr 
Total Tracers  & 43.1 & 41.0    \cr  \\ 
\enddata
\label{table:sequels_eboss_riz_req}
\end{deluxetable}

\begin{figure}[btp]
\centering
\includegraphics[scale=0.59]{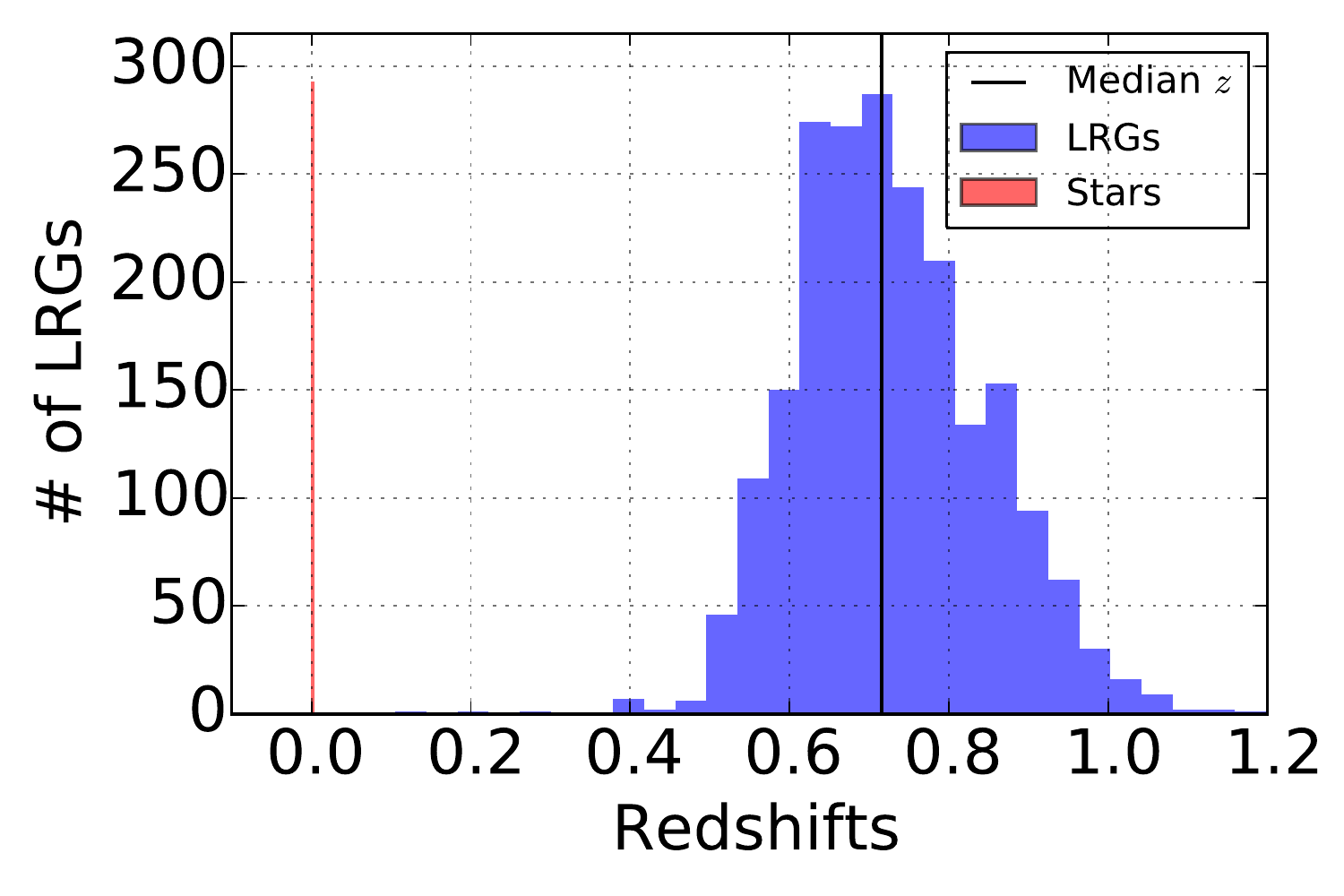}
	\caption{ {Redshift histogram of 2,119 visually inspected LRGs (blue bar) observed with \eboss. The median redshift of confirmed galaxies is 0.712 (black line), with 9$\%$ stellar contamination (red bar). We use only objects with secure redshifts ($\texttt{z\_conf} > 1$) here.}}
\label{fig:riw_sequels_redshift}
\end{figure}

\begin{figure*}[btp]
\centering
\includegraphics[scale=0.95]{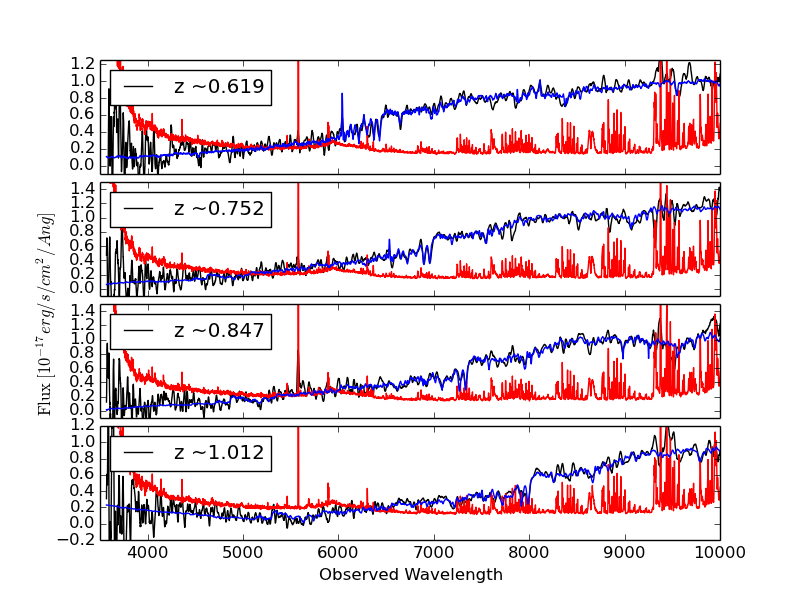}
	\caption{ {Representative spectra of galaxies from the \eboss\ LRG sample, smoothed with a 21 pixel boxcar kernel. Shown are four LRGs covering the entire redshift regime of $0.6~\leqsim\,z~\leqsim\,1$. Flux errors are plotted in red while the template model fits are in blue. Black curves depict the observed spectra.}}
\label{fig:sequels_spectra}
\end{figure*}

In Figure~\ref{fig:riw_sequels_redshift}, we present the overall redshift distribution ($N(z)$) of the visually-inspected \eboss\ LRGs. Although we fail to meet the requirement of $80\%$ efficiency at targeting $0.6 < z < 1.0$ LRGs, our target selection algorithm still exceeds the median redshift requirement, which is calculated only for actual galaxies (and hence includes only non-stellar targets with robust redshift measurements). In Figure~\ref{fig:sequels_spectra}, we show examples of LRG spectra across the redshift range of interest for \eboss. There is an excellent match between the measured SEDs and the templates, confirming the robustness  of these redshift measurements. The \eboss\ LRG sample can be augmented with $z > 0.6 $ \boss\ CMASS LRGs to meet our  requirements on the total number of LRG redshifts within the range $0.6 < z < 1.0$; as a result, we still expect to achieve a $1\%$ measurement of the LRG BAO scale at $z \sim 0.7$, even though the LRG sample falls short of its requirements. 
 
\section{Tests of homogeneity and implications for large-scale clustering measurements}\label{sec:uni_test}
As discussed in \S\ref{sec:lrg_eboss_goals}, we require that the target sample be highly uniform to prevent non-cosmological signals from contaminating clustering measurements.  Exploring systematics that can affect the inferred clustering of targets is often considered only when survey data is used for science analyses. 
We instead have investigated these issues while exploring target selection methods, enabling more informed decisions regarding survey strategy. For instance, foreknowledge of which areas of the survey may pose problems for controlling clustering measurements potentially allows the survey footprint to be shifted.

We assess the uniformity of the target sample by comparing the observed density of targets to maps of local imaging conditions and Galactic structure. We apply a regression analysis of surface density against a broad set of tracers of potential systematics; the intention is similar to, e.g., \cite{Scranton02,Ross11,Ho12, Leistedt13,Gian14}, but unlike those works, we simultaneously fit for the impact of a wide variety of systematics rather than correlating against one at a time.  This has the advantage of producing a model of systematic-affected density that will provide accurate predictions for the combined effects of all the systematics considered, even if the input systematic maps are covariant with each other (as, for instance, stellar density and dust extinction must inevitably be).   

We focus on systematics associated with imaging data characteristics or with known astrophysical effects such as dust extinction and stellar density.  Using the results of the regression analysis (described below) we assemble maps of the observed density and the predicted density.  We identify regions within our footprint where the total span of target density fluctuation is less than 15\%, and consider the portion of sky with larger variations to be contaminated at an unacceptable level; this criterion is based on prior experience with the level of systematics that may be corrected reliably in BOSS \cite{Ross11}.  We note that fluctuations in density within the final $0.6 < z < 1.0$ LRG catalog are likely to be smaller than this, as once spectra are obtained, stars and redshift outliers can be removed; such objects are naturally expected to be less homogenous over the SDSS survey area than the true LRGs.

\subsection{Homogeneity of \eboss\ LRG targets}\label{sec:regress_test}
To begin, we identify a broad set of imaging parameters that could affect \eboss\ target selection:

\begin{enumerate}
\item \textbf{\tt W1covmedian}: The median number of single-exposure frames per pixel in the WISE W1 band.

\item  \textbf{\tt moon\_lev}: The fraction of frames that were contaminated with scattered moon light in the WISE W1 band.

\item  \textbf{\tt W1median}: The median of accumulated flux per pixel in the WISE W1 band measured in units of $DN$ (data number).\footnote{ The accumulated photons in each pixel are represented by a number in units of $DN$ }.

\item \textbf{Galactic Latitude}: used as a proxy for stellar contamination.

\item \textbf{Galactic extinction}: We use \textit{r} band extinction, as given by SFD.

\item \textbf{FWHM} in the \sdss \textit{z}-band: We use FWHM as an estimate of the 'seeing' or imaging quality for the SDSS imaging.

\item \textbf{SKYFLUX} in the \sdss \textit{z}-band: the background sky level affects the detection of faint objects is more difficult in the brighter regions of the sky .
\end{enumerate}
We create maps of the WISE systematics over the entire \sdss\ footprint using the metadata tables associated with the Atlas images and source tables provided by WISE survey team; {\tt W1covmedian}, {\tt W1median}, and {\tt moon\_lev} are all quantities in these tables.\footnote{\url{http://wise2.ipac.caltech.edu/docs/release/allsky/expsup\\/sec2\_4f.html}} We use the seeing and the sky-background in the \textit{z} band since the \eboss\ LRG selection algorithm is flux-limited in that bandpass filter.  Due to the scan strategy of SDSS, the seeing and sky background in other SDSS bands should correlate strongly with this quantity, making the use of multiple filters' quantities redundant. 

Next, we break the sky up into equal-area pixels of 0.36 \degsq\ and weight all pixels equally. The observed density, $SD_{obs}$, in each pixel can be expressed as a combination of a mean level, the impact of all of the systematics, and random noise:

\begin{align}
      SD_{obs} = S_{0} + \sum\limits_{i=1}^7 S_{i} \times x_{i} + \epsilon,\label{eqn:sd_eqn}    
\end{align}
where $S_{0}$ is the constant term representing the mean density of objects in each pixel, $S_{i}$ are the coefficients for the values of each individual source of potential systematics fluctuations in that pixel ($x_{i}$), and $\epsilon$ represents the combined effect of Poisson noise (or shot noise) and sample/cosmic variance in that pixel. For larger pixels such that the mean pixel target density is $\sim$15 or more, the Poisson noise can be approximated as a Gaussian. Under these conditions, multi-linear regression provides an effective means of determining the unknown coefficients, $S_{0}$ and $S_{i}$.   We derive a best-fit model based on 
 minimizing the value of reduced-$\chi^{2}$ ($\chi^{2}$ per degree of freedom).  We have explored larger or smaller pixelizations and find that our results are unchanged.

The coefficients obtained from this multi-linear regression are then used in combination with the maps of potential systematics to predict the target density across the whole footprint, producing a
statistic that we will refer to as the {\em Predicted Surface Density} or $PSD$. We also define a {\em Residual Surface Density}, or $Residual\_SD_j$, for any particular systematic as the difference between $SD_{obs}$ and the $Reduced\_PSD_j$ (which is calculated by omitting the $j$'th systematic term in calculating the PSD). This quantity should be linear in systematic $j$ with a slope corresponding to $S_j$ if our linear regression model is appropriate to the problem. To summarize our formalism:

\begin{align}
      PSD = S_{0} + \sum\limits_{i=1}^n S_{i} \times x_{i},\label{eqn:psd_eqn}\\
      Reduced\_PSD_j = PSD - S_{j} \times x_{j}, {\rm and}\label{eqn:rpsd_eqn}\\
      Residual\_SD_j =  SD_{obs} - (PSD - S_{j} \times x_{j}) ,\label{eqn:rsd_eqn}.    
\end{align}

\indent where the $j$ index indicates a single systematic of interest. 

\subsection{Predicted Surface Density for \eboss\ LRG targets}\label{sec:psd}
The $PSD$ is highly useful for testing the uniformity of the target sample across the whole footprint, enabling comparisons to survey requirements. We find that the effects of systematics
produce significantly different best-fit models (in terms of both the mean density and the coefficients for each systematic) in the areas of \sdss\ imaging around the Northern Galactic Cap (NGC) and the Southern Galactic Cap (SGC). However, for both the regions considered independently, multi-linear regression provides an acceptable best-fit model. Hence we analyze these regions separately. 

\begin{figure*}[btp]
\centering
\includegraphics[scale=2.3]{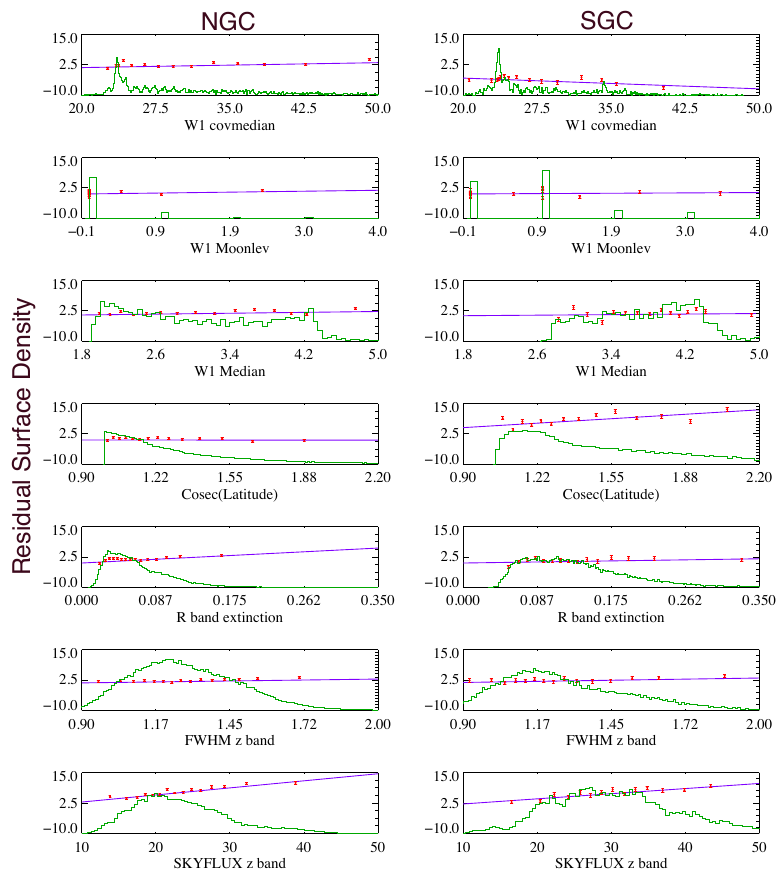}
	\caption{ {The residual surface density from the regression model with a single systematic omitted as a function of the systematic left out, along with the corresponding predictions from the regression model, for all systematic maps considered. Individual points have been averaged over 4000 sky pixels in the Northern Galactic Cap (NGC) or 2000 pixels in the Southern Galactic Cap (SGC). The straight line shows the prediction of the regression model for the impact of the systematic indicated on the x-axis (cf. equation~\ref{eqn:sd_eqn}).  The overplotted histograms show the distribution of pixel values, and correspond to the y axis at the right side of each plot. The left-hand column of plots are for pixels in the NGC, while the right-hand column of plots are for the SGC.  A linear model appears to be appropriate for all systematics considered.}}
\label{fig:lrg_regress_mp}
\end{figure*}

The resulting regression 
fits are shown in Figure~\ref{fig:lrg_regress_mp}. In these plots, we plot the $Residual\_SD$ for each individual systematic which was been left out in calculating $Residual\_SD_j$. The data points plotted are averages over 4000 sky pixels in the NGC or 2000 sky-pixels in the SGC; the error bars represent the standard error on the mean for each point. The straight lines represent the prediction from the regression model for the impact of the systematic indicated on the $x$-axis, $x_j$ (cf. Eqn.~\ref{eqn:sd_eqn}); i.e., we plot $y = S_j \times x_j$.

\subsection{Analysis of regression results}\label{sec:summary_regress}

\begin{figure}[btp]
\centering
\includegraphics[scale=0.40]{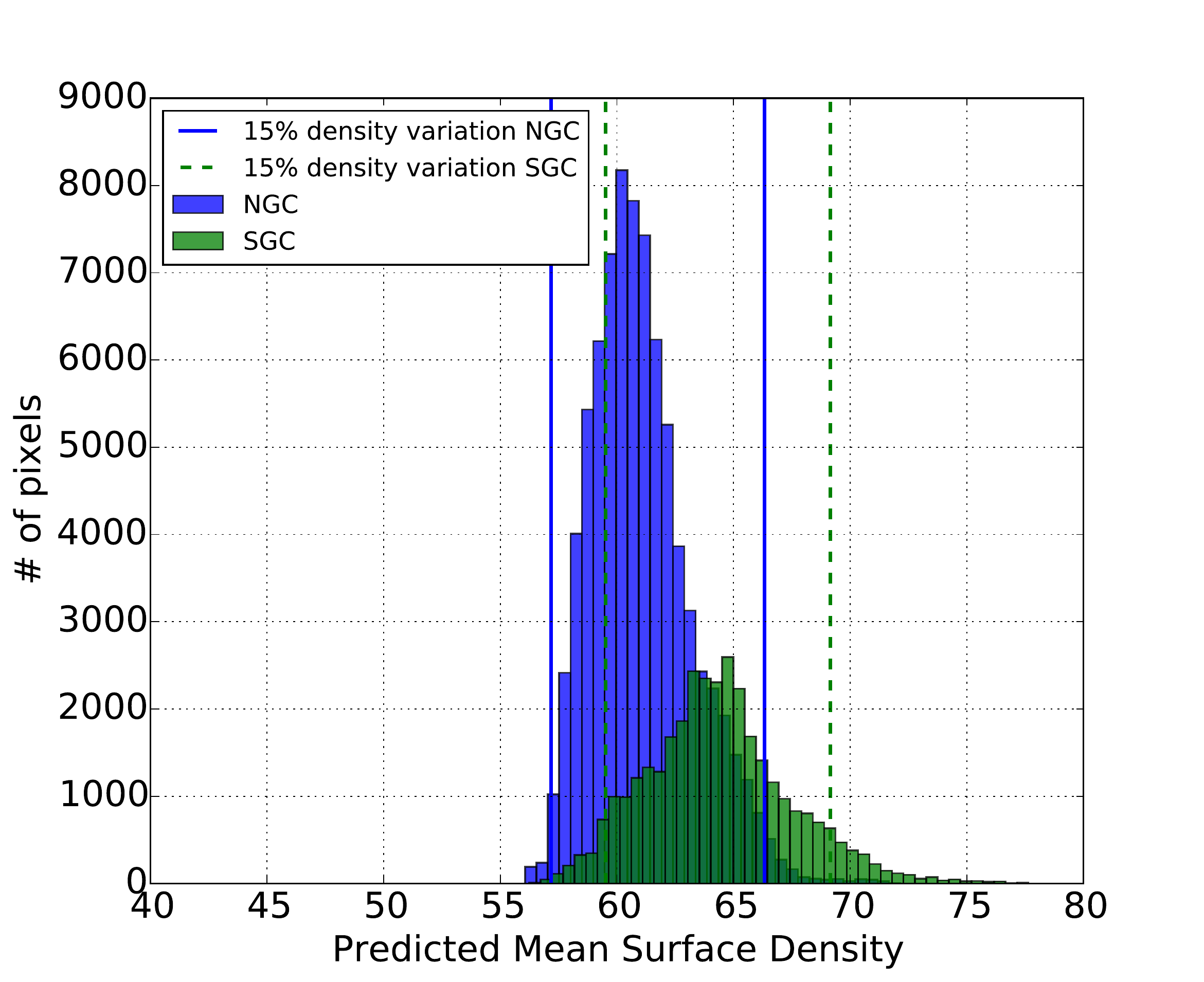}
	\caption{ {Histogram of the surface density predicted by the regression models described in \S\ref{sec:regress_test}. The blue bars represent the NGC, with solid blue lines depicting the $15\%$ window within which samples are expected to be sufficiently homogeneous for robust large-scale-structure measurements. Similarly, the green bars represent the density in the SGC, with dotted green lines depicting the $15\%$ window. We find that $\sim97\%$ of the NGC footprint with \sdss\ imaging meets the homogeneity requirements of \eboss\ (see \S\ref{sec:ts_req}). However, in the SGC, only $\sim82\%$ of the possible \eboss\ footprint meets these requirements.}}
\label{fig:lrg_regress_pd}
\end{figure}

Our regression analysis allows us to determine what fraction of the survey footprint satisfies the requirement of less than $15\%$ total variation in target density (point 6 in \S\ref{sec:ts_req}). This $15\%$ window is not necessarily symmetric around the mean, so we fix its limits such that the footprint area satisfying the requirement 
is maximized. The windows containing regions with $PSD$ variation $< 15\%$ are overplotted on the histograms of predicted density in Figure~\ref{fig:lrg_regress_pd}. In the NGC, $\sim97\%$ of the imaging area meets the \eboss\ survey requirements for homogeneity. However, in the SGC, 
only $\sim82\%$ of the area meets these requirements. At worst, these fluctuations will require that 8\% of the total 7500\degsq\ eBOSS area is masked. However, these fluctuations may be reduced once spectroscopic redshifts are obtained; we will perform a similar analysis on the final spectroscopic sample in later work.

Differences in the observed number density between the NGC and SGC were found for the \boss\ \texttt{CMASS} and \texttt{LOWZ} samples, and were analyzed in depth by \cite{ashley2012}. These differences matched the photometric offsets between the two regions determined by \citet{snf_sdss2011}. These offsets have been incorporated into the re-calibrated photometry used for \eboss; any difference in target density between the regions is therefore due to still-unknown differences between the two regions. This issue will require further investigation in future \eboss~ studies.

Based on the regression model, we can assess which systematics are most strongly affecting target selection. We find that all of the potential WISE imaging systematics have relatively weak effects on the density of selected targets. This can be seen from the flatness of $Residual\_SD$ for these parameters in Figure~\ref{fig:lrg_regress_mp}. The most significant effects are associated with dust extinction, stellar contamination, and the \sdss\ sky background level, as seen from the steep slopes in Figure~\ref{fig:lrg_regress_mp}.  It is unclear whether dust or stellar contamination is more fundamentally responsible for variations in density, since the two correlate with each other strongly.  Given the variation in coefficients, it is likely that the same phenomenon is being ascribed more to dust in the NGC and to Galactic latitude in the SGC, and those differences in coefficient are not truly significant.  Fortunately, the regression model will still predict the correct density from covariant variables such as these, regardless of which covariate is actually responsible.  

\begin{figure*}[btp]
\centering
	\includegraphics[scale=0.46]{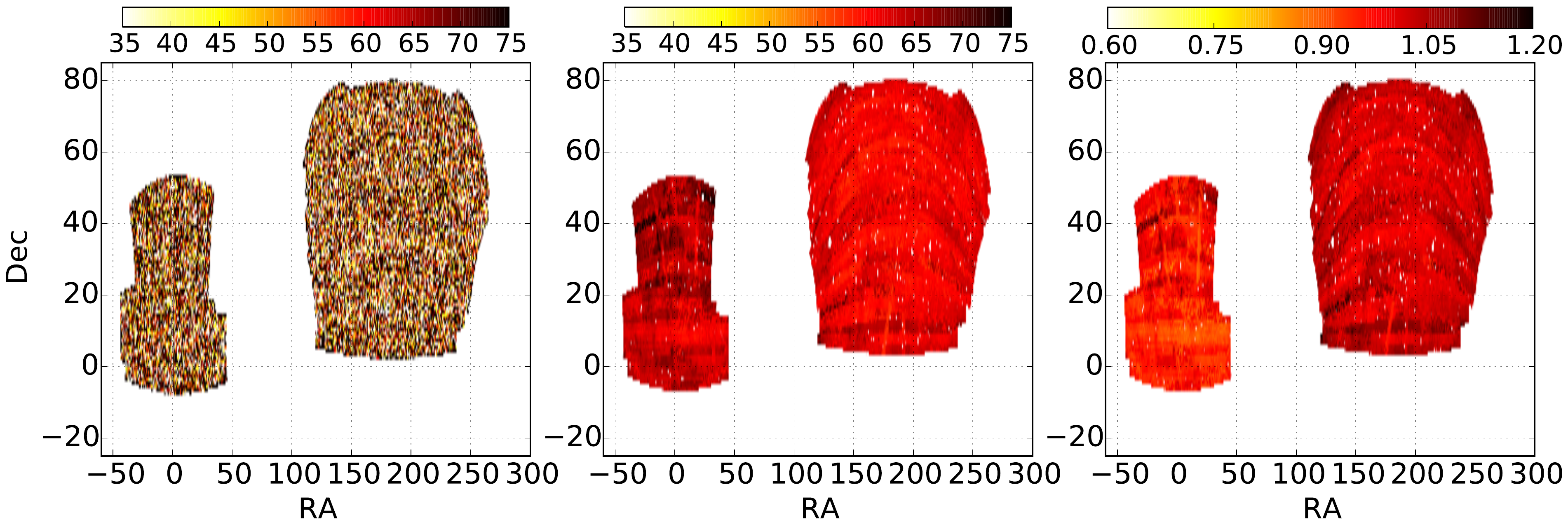}
	\caption{{Observed surface density map of \eboss\ LRGs over the area of the \sdss\ imaging footprint used to derive targets for the \boss\ survey. \eboss\ will target more than $\sim$375{,}000 LRGs over a $\sim$7500 \degsq\ subset of this area, corresponding to a surface density of $\sim$50 \perdegsq. The color scale in this panel is dominated by Poisson noise and sample/cosmic variance.  The middle panel shows a similar plot based on the predicted density, while the third panel shows the regions which would be masked to reach \eboss\ homogeneity requirements. Pixels shown with a shade that is fainter or darker than the typical are those which fail to meet the $15\%$ target density variation requirement of \S\ref{sec:ts_req}.}}
\label{fig:lrg_skymap}
\end{figure*}

We depict the observed surface density, the predicted surface density, and the mask of the survey across the whole footprint of \sdss\ in Figure~\ref{fig:lrg_skymap}. 

\subsection{Impact of zero point variations}\label{sec:zero_point}
 We next assess the expected level of variation in target density due to errors in zero-point calibrations, which can then be compared to the targeting requirements.  We investigate this by determining the fractional derivative in the number of targets selected ($N$) as we shift all magnitudes in a given band ($m$) by a constant amount -- i.e., we calculate ${1\over N}dN/dm$ -- and then assess what impact this sensitivity has on target density. We find that zero-point errors of 0.01 magnitude in the \textit{r}, \textit{i}, \textit{z}, and \textit{W1} bands causes fractional changes of 2.26\%, 2.5\%, 6.24\%, and 0.6\%, respectively, in the target density of the LRG sample. Finkbeiner et al. (2014) estimate that the 1$\sigma$ zero point uncertainties ($\sigma_{zp}$) after recalibration of SDSS are  7, 7, and 8 millimagnitudes in the \sdss\ \textit{r}, \textit{i}, and \textit{z} bands respectively, while {\it WISE} calibration uncertainties in the W1 band are approximately 0.016 mag \citep{wise_sup_jarret_2011}. 
 
\begin{figure}[btp]
\centering
\includegraphics[scale=0.59]{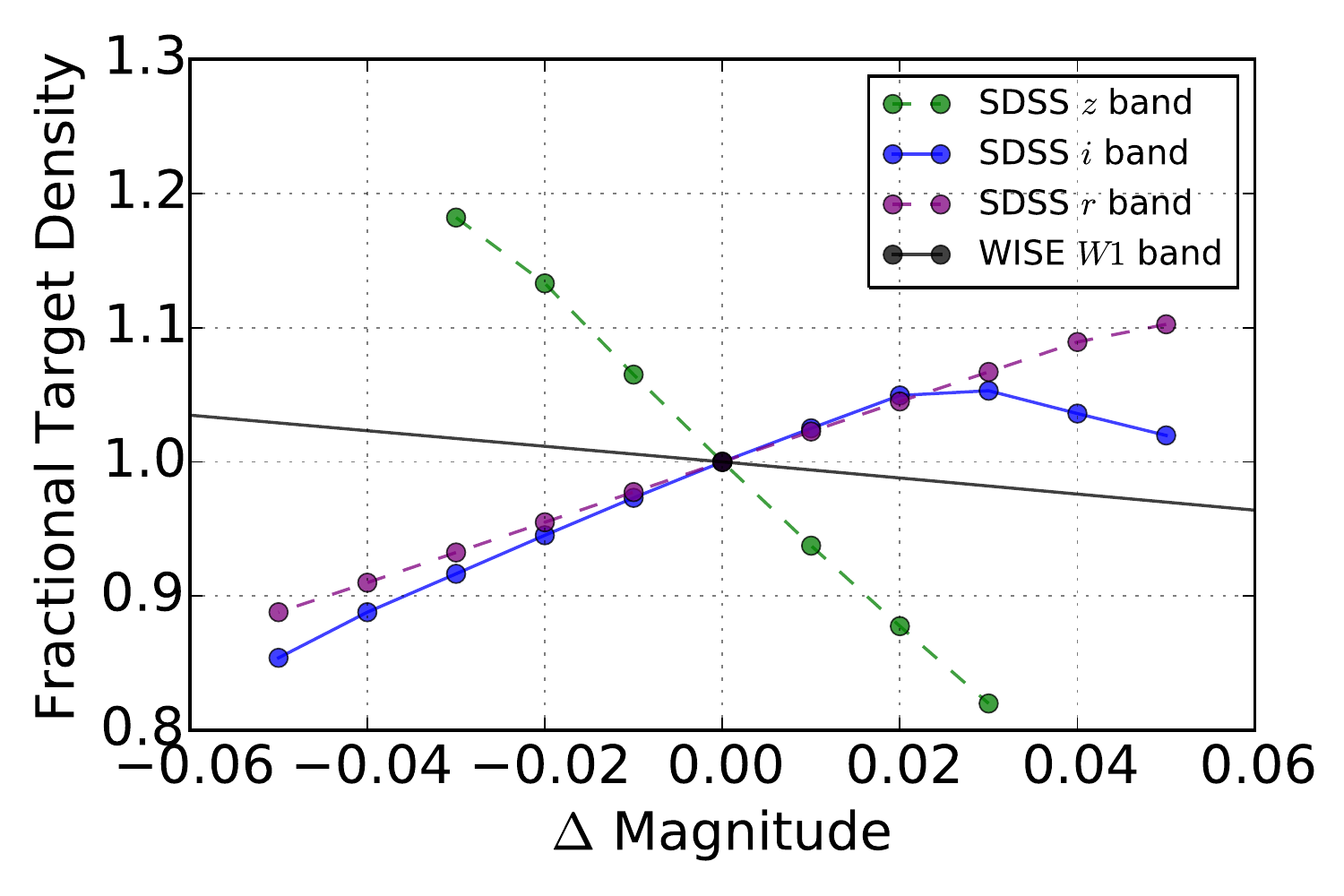}
	\caption{ {Change in target density as a function of an overall shift in all magnitudes in either the \sdss\ \textit{r}, \textit{i}, \textit{z}, or the \wise\ \textit{W1} band. Given the current level of zero point uncertainty in SDSS and \wise\ photometry, the LRG target selection is only sensitive to the uncertainty in the zero point of the \textit{z} band.}}
\label{fig:lrg_zero_point}
\end{figure}

Assuming that zero point errors will be Gaussian-distributed, 95\% of all points on the sky will be within $\pm$2$\sigma$ of the mean zero point.  Hence, the total fractional variation in density over that area will be ${4\over N}\times \mid \Delta N/ \Delta m\mid \times \sigma_{zp}$.  We present the results of this calculation in Table~\ref{table:zero_point} and Figure~\ref{fig:lrg_zero_point}.  For all bands but \textit{z}, the impact of zero point variations on the density of LRG targets will be minimal.  However, the estimated level of \textit{z} band zero point uncertainty is sufficiently large that more than 13\% of the \eboss\ area will go beyond the 15\% target density variation requirement.  The impact of this variation and strategies for mitigating it will be explored in future \eboss\ papers.\\
\begin{deluxetable*}{llcrr}
\tablecaption{Summary of variations in target density due to errors in \sdss\ imaging zero points.}
\tablehead{ Bands & Derivative of &RMS zero-point & 95\% range of  variation \cr 
 &fractional density & error & in fractional density \cr
  &\textbf{$\mid {1\over N}\Delta N/ \Delta m\mid$} &($\sigma_{zp}$) & $4\times \mid \Delta \bar N/ \Delta Mag\mid \times \sigma_{zp}$}
\startdata
$\sdss\ r$& 2.26  & $7\times10^{-3}$ & 0.063 \cr
$\sdss\ i$& 2.5 &$7\times10^{-3} $&0.070 \cr
$\sdss\ z$ & 6.24 & $8\times10^{-3}$& 0.199  \cr
$WISE\ W1$ & 0.60 & $16\times10^{-3}$& 0.038 \cr
\enddata
\tablecomments{The impact of zero point uncertainties on the density of targets. The \eboss\ LRG sample meets the requirement that density variations due to zero-point errors be less than 15\% in the \sdss\ \textit{r} and \textit{i} bands, but fails to meet that criterion in the \textit{z} band with current calibrations.}
\label{table:zero_point}
\end{deluxetable*}

\section{Conclusions}\label{sec:con}

The LRG component of \sdssiv/\eboss\ will obtain spectroscopy of a sample of over 375{,}000 potential intrinsically luminous early-type galaxies at $z > 0.6$. Based on the initial set of \eboss\ data, we find that the efficiency of this selection for selecting $0.6 < z < 1.0$ LRGs with secure redshift measurementsis $\sim 68-72\%$. This is lower than our requirement (80\%). In addition, 9\% of stellar contamination. The sample is flux-limited to keep the selection algorithm robust, as well as to maintain a sufficient signal-to-noise ratio to enable the resulting LRG spectra to provide secure redshift measurements. We have made some minor improvements to our selection algorithms post-\sequels\ which should further improve the performance of our LRG-targeting algorithm. \eboss\ LRGs will be augmented by $z > 0.6$ \boss\ LRGs to achieve the science goals of \eboss.\\
\indent The primary science drivers of the \eboss\ LRG sample are to study the large-scale structure of the Universe out to $z \sim 1$.  With careful control of incompletenesses and selection effects, the \eboss\ LRG algorithm will also provide a large sample for galaxy evolution studies of giant elliptical galaxies. 
The \sdssiv/\eboss\ LRGs will cover a volume either not probed, or not probed at high density, by \sdssiii/\boss, and will allow enable both BAO and RSD measurements with a highly uniform set of luminous, early-type galaxies. The \sdssiv/\eboss\ LRG sample will provide
a powerful extension of \sdssiii/\boss\ for the study of structure and galaxy evolution at high redshifts.\\

\acknowledgements This work was supported by an Early Career Grant from the U.S. Department of Energy Office of Science.  
This paper represents an effort by both the SDSS-III and SDSS-IV collaborations.
Funding for SDSS-III was provided by the Alfred
P. Sloan Foundation, the Participating Institutions, the
National Science Foundation, and the U.S. Department
of Energy Office of Science.

This paper includes targets derived from the images of
the Wide-Field Infrared Survey Explorer, which is a
joint project of the University of California, Los Angeles,
and the Jet Propulsion Laboratory/California Institute
of Technology, funded by the National Aeronautics and
Space Administration.

Funding for the Sloan Digital Sky Survey IV has been provided by
the Alfred P. Sloan Foundation, the U.S. Department of Energy Office of
Science, and the Participating Institutions. SDSS-IV acknowledges
support and resources from the Center for High-Performance Computing at
the University of Utah. The SDSS web site is www.sdss.org.

SDSS-IV is managed by the Astrophysical Research Consortium for the
Participating Institutions of the SDSS Collaboration including the
Brazilian Participation Group, the Carnegie Institution for Science,
Carnegie Mellon University, the Chilean Participation Group,
the French Participation Group, Harvard-Smithsonian Center for Astrophysics,
Instituto de Astrof\'isica de Canarias, The Johns Hopkins University,
Kavli Institute for the Physics and Mathematics of the Universe (IPMU) /
University of Tokyo, Lawrence Berkeley National Laboratory,
Leibniz Institut f\"ur Astrophysik Potsdam (AIP),
Max-Planck-Institut f\"ur Astronomie (MPIA Heidelberg),
Max-Planck-Institut f\"ur Astrophysik (MPA Garching),
Max-Planck-Institut f\"ur Extraterrestrische Physik (MPE),
National Astronomical Observatory of China, New Mexico State University,
New York University, University of Notre Dame,
Observat\'ario Nacional / MCTI, The Ohio State University,
Pennsylvania State University, Shanghai Astronomical Observatory,
United Kingdom Participation Group,
Universidad Nacional Aut\'onoma de M\'exico, University of Arizona,
University of Colorado Boulder, University of Portsmouth,
University of Utah, University of Virginia, University of Washington,
University of Wisconsin,
Vanderbilt University, and Yale University.

\bibliographystyle{apj}
\bibliography{lrg_ref}\scriptsize

\begin{thebibliography}{}
\expandafter\ifx\csname natexlab\endcsname\relax\def\natexlab#1{#1}\fi

\bibitem[{{Abazajian} {et~al.}(2009){Abazajian}, {Adelman-McCarthy},
  {Ag{\"u}eros}, {Allam}, {Allende Prieto}, {An}, {Anderson}, {Anderson},
  {Annis}, {Bahcall}, \& et~al.}]{DR7}
{Abazajian}, K.~N., {Adelman-McCarthy}, J.~K., {Ag{\"u}eros}, M.~A., {et~al.}
  2009, \apjs, 182, 543

\bibitem[{{Aihara} {et~al.}(2011){Aihara}, {Allende Prieto}, {An}, {Anderson},
  {Aubourg}, {Balbinot}, {Beers}, {Berlind}, {Bickerton}, {Bizyaev}, {Blanton},
  {Bochanski}, {Bolton}, {Bovy}, {Brandt}, {Brinkmann}, {Brown}, {Brownstein},
  {Busca}, {Campbell}, {Carr}, {Chen}, {Chiappini}, {Comparat}, {Connolly},
  {Cortes}, {Croft}, {Cuesta}, {da Costa}, {Davenport}, {Dawson}, {Dhital},
  {Ealet}, {Ebelke}, {Edmondson}, {Eisenstein}, {Escoffier}, {Esposito},
  {Evans}, {Fan}, {Femen{\'{\i}}a Castell{\'a}}, {Font-Ribera}, {Frinchaboy},
  {Ge}, {Gillespie}, {Gilmore}, {Gonz{\'a}lez Hern{\'a}ndez}, {Gott}, {Gould},
  {Grebel}, {Gunn}, {Hamilton}, {Harding}, {Harris}, {Hawley}, {Hearty}, {Ho},
  {Hogg}, {Holtzman}, {Honscheid}, {Inada}, {Ivans}, {Jiang}, {Johnson},
  {Jordan}, {Jordan}, {Kazin}, {Kirkby}, {Klaene}, {Knapp}, {Kneib},
  {Kochanek}, {Koesterke}, {Kollmeier}, {Kron}, {Lampeitl}, {Lang}, {Le Goff},
  {Lee}, {Lin}, {Long}, {Loomis}, {Lucatello}, {Lundgren}, {Lupton}, {Ma},
  {MacDonald}, {Mahadevan}, {Maia}, {Makler}, {Malanushenko}, {Malanushenko},
  {Mandelbaum}, {Maraston}, {Margala}, {Masters}, {McBride}, {McGehee},
  {McGreer}, {M{\'e}nard}, {Miralda-Escud{\'e}}, {Morrison}, {Mullally},
  {Muna}, {Munn}, {Murayama}, {Myers}, {Naugle}, {Neto}, {Nguyen}, {Nichol},
  {O'Connell}, {Ogando}, {Olmstead}, {Oravetz}, {Padmanabhan},
  {Palanque-Delabrouille}, {Pan}, {Pandey}, {P{\^a}ris}, {Percival},
  {Petitjean}, {Pfaffenberger}, {Pforr}, {Phleps}, {Pichon}, {Pieri}, {Prada},
  {Price-Whelan}, {Raddick}, {Ramos}, {Reyl{\'e}}, {Rich}, {Richards}, {Rix},
  {Robin}, {Rocha-Pinto}, {Rockosi}, {Roe}, {Rollinde}, {Ross}, {Ross},
  {Rossetto}, {S{\'a}nchez}, {Sayres}, {Schlegel}, {Schlesinger}, {Schmidt},
  {Schneider}, {Sheldon}, {Shu}, {Simmerer}, {Simmons}, {Sivarani}, {Snedden},
  {Sobeck}, {Steinmetz}, {Strauss}, {Szalay}, {Tanaka}, {Thakar}, {Thomas},
  {Tinker}, {Tofflemire}, {Tojeiro}, {Tremonti}, {Vandenberg}, {Vargas
  Maga{\~n}a}, {Verde}, {Vogt}, {Wake}, {Wang}, {Weaver}, {Weinberg}, {White},
  {White}, {Yanny}, {Yasuda}, {Yeche}, \& {Zehavi}}]{DR8}
{Aihara}, H., {Allende Prieto}, C., {An}, D., {et~al.} 2011, \apjs, 195, 26

\bibitem[{{Alam} {et~al.}(2015){Alam}, {Albareti}, {Allende Prieto}, {Anders},
  {Anderson}, {Anderton}, {Andrews}, {Armengaud}, {Aubourg}, {Bailey}, \&
  et~al.}]{DR12}
{Alam}, S., {Albareti}, F.~D., {Allende Prieto}, C., {et~al.} 2015, \apjs, 219,
  12

\bibitem[{{Beutler} {et~al.}(2014{\natexlab{a}}){Beutler}, {Saito},
  {Brownstein}, {Chuang}, {Cuesta}, {Percival}, {Ross}, {Ross}, {Schneider},
  {Samushia}, {S{\'a}nchez}, {Seo}, {Tinker}, {Wagner}, \&
  {Weaver}}]{Beutler14a}
{Beutler}, F., {Saito}, S., {Brownstein}, J.~R., {et~al.} 2014{\natexlab{a}},
  \mnras, 444, 3501

\bibitem[{{Beutler} {et~al.}(2014{\natexlab{b}}){Beutler}, {Saito}, {Seo},
  {Brinkmann}, {Dawson}, {Eisenstein}, {Font-Ribera}, {Ho}, {McBride},
  {Montesano}, {Percival}, {Ross}, {Ross}, {Samushia}, {Schlegel},
  {S{\'a}nchez}, {Tinker}, \& {Weaver}}]{Beutler14b}
{Beutler}, F., {Saito}, S., {Seo}, H.-J., {et~al.} 2014{\natexlab{b}}, \mnras,
  443, 1065

\bibitem[{{Bolton} {et~al.}(2012){Bolton}, {Schlegel}, {Aubourg}, {Bailey},
  {Bhardwaj}, {Brownstein}, {Burles}, {Chen}, {Dawson}, {Eisenstein}, {Gunn},
  {Knapp}, {Loomis}, {Lupton}, {Maraston}, {Muna}, {Myers}, {Olmstead},
  {Padmanabhan}, {P{\^a}ris}, {Percival}, {Petitjean}, {Rockosi}, {Ross},
  {Schneider}, {Shu}, {Strauss}, {Thomas}, {Tremonti}, {Wake}, {Weaver}, \&
  {Wood-Vasey}}]{bolton2012}
{Bolton}, A.~S., {Schlegel}, D.~J., {Aubourg}, {\'E}., {et~al.} 2012, \aj, 144,
  144

\bibitem[{{Cannon} {et~al.}(2006){Cannon}, {Drinkwater}, {Edge}, {Eisenstein},
  {Nichol}, {Outram}, {Pimbblet}, {de Propris}, {Roseboom}, {Wake}, {Allen},
  {Bland-Hawthorn}, {Bridges}, {Carson}, {Chiu}, {Colless}, {Couch}, {Croom},
  {Driver}, {Fine}, {Hewett}, {Loveday}, {Ross}, {Sadler}, {Shanks}, {Sharp},
  {Smith}, {Stoughton}, {Weilbacher}, {Brunner}, {Meiksin}, \&
  {Schneider}}]{Cannon2006}
{Cannon}, R., {Drinkwater}, M., {Edge}, A., {et~al.} 2006, \mnras, 372, 425

\bibitem[{{Comparat} {et~al.}(2015){Comparat}, {et al.}, \& {\em in
  preparation}}]{Comparat_elg}
{Comparat}, J., {et al.}, \& {\em in preparation}. 2015, to be submitted to
  \apj

\bibitem[{{Dawson} {et~al.}(2015){Dawson}, {et al.}, \& {\em in
  preparation}}]{eboss_overview}
{Dawson}, K., {et al.}, \& {\em in preparation}. 2015, to be submitted to \apj

\bibitem[{{Dawson} {et~al.}(2013){Dawson}, {Schlegel}, {Ahn}, {Anderson},
  {Aubourg}, {Bailey}, {Barkhouser}, {Bautista}, {Beifiori}, {Berlind},
  {Bhardwaj}, {Bizyaev}, {Blake}, {Blanton}, {Blomqvist}, {Bolton}, {Borde},
  {Bovy}, {Brandt}, {Brewington}, {Brinkmann}, {Brown}, {Brownstein}, {Bundy},
  {Busca}, {Carithers}, {Carnero}, {Carr}, {Chen}, {Comparat}, {Connolly},
  {Cope}, {Croft}, {Cuesta}, {da Costa}, {Davenport}, {Delubac}, {de Putter},
  {Dhital}, {Ealet}, {Ebelke}, {Eisenstein}, {Escoffier}, {Fan}, {Filiz Ak},
  {Finley}, {Font-Ribera}, {G{\'e}nova-Santos}, {Gunn}, {Guo}, {Haggard},
  {Hall}, {Hamilton}, {Harris}, {Harris}, {Ho}, {Hogg}, {Holder}, {Honscheid},
  {Huehnerhoff}, {Jordan}, {Jordan}, {Kauffmann}, {Kazin}, {Kirkby}, {Klaene},
  {Kneib}, {Le Goff}, {Lee}, {Long}, {Loomis}, {Lundgren}, {Lupton}, {Maia},
  {Makler}, {Malanushenko}, {Malanushenko}, {Mandelbaum}, {Manera}, {Maraston},
  {Margala}, {Masters}, {McBride}, {McDonald}, {McGreer}, {McMahon}, {Mena},
  {Miralda-Escud{\'e}}, {Montero-Dorta}, {Montesano}, {Muna}, {Myers},
  {Naugle}, {Nichol}, {Noterdaeme}, {Nuza}, {Olmstead}, {Oravetz}, {Oravetz},
  {Owen}, {Padmanabhan}, {Palanque-Delabrouille}, {Pan}, {Parejko},
  {P{\^a}ris}, {Percival}, {P{\'e}rez-Fournon}, {P{\'e}rez-R{\`a}fols},
  {Petitjean}, {Pfaffenberger}, {Pforr}, {Pieri}, {Prada}, {Price-Whelan},
  {Raddick}, {Rebolo}, {Rich}, {Richards}, {Rockosi}, {Roe}, {Ross}, {Ross},
  {Rossi}, {Rubi{\~n}o-Martin}, {Samushia}, {S{\'a}nchez}, {Sayres}, {Schmidt},
  {Schneider}, {Sc{\'o}ccola}, {Seo}, {Shelden}, {Sheldon}, {Shen}, {Shu},
  {Slosar}, {Smee}, {Snedden}, {Stauffer}, {Steele}, {Strauss}, {Streblyanska},
  {Suzuki}, {Swanson}, {Tal}, {Tanaka}, {Thomas}, {Tinker}, {Tojeiro},
  {Tremonti}, {Vargas Maga{\~n}a}, {Verde}, {Viel}, {Wake}, {Watson}, {Weaver},
  {Weinberg}, {Weiner}, {West}, {White}, {Wood-Vasey}, {Yeche}, {Zehavi},
  {Zhao}, \& {Zheng}}]{dawson_boss_2013}
{Dawson}, K.~S., {Schlegel}, D.~J., {Ahn}, C.~P., {et~al.} 2013, \aj, 145, 10

\bibitem[{{Eisenstein} {et~al.}(2001){Eisenstein}, {Annis}, {Gunn}, {Szalay},
  {Connolly}, {Nichol}, {Bahcall}, {Bernardi}, {Burles}, {Castander},
  {Fukugita}, {Hogg}, {Ivezi{\'c}}, {Knapp}, {Lupton}, {Narayanan}, {Postman},
  {Reichart}, {Richmond}, {Schneider}, {Schlegel}, {Strauss}, {SubbaRao},
  {Tucker}, {Vanden Berk}, {Vogeley}, {Weinberg}, \& {Yanny}}]{Eisenstein2001}
{Eisenstein}, D.~J., {Annis}, J., {Gunn}, J.~E., {et~al.} 2001, \aj, 122, 2267

\bibitem[{{Eisenstein} {et~al.}(2005){Eisenstein}, {Zehavi}, {Hogg},
  {Scoccimarro}, {Blanton}, {Nichol}, {Scranton}, {Seo}, {Tegmark}, {Zheng},
  {Anderson}, {Annis}, {Bahcall}, {Brinkmann}, {Burles}, {Castander},
  {Connolly}, {Csabai}, {Doi}, {Fukugita}, {Frieman}, {Glazebrook}, {Gunn},
  {Hendry}, {Hennessy}, {Ivezi{\'c}}, {Kent}, {Knapp}, {Lin}, {Loh}, {Lupton},
  {Margon}, {McKay}, {Meiksin}, {Munn}, {Pope}, {Richmond}, {Schlegel},
  {Schneider}, {Shimasaku}, {Stoughton}, {Strauss}, {SubbaRao}, {Szalay},
  {Szapudi}, {Tucker}, {Yanny}, \& {York}}]{Eisenstein2005}
{Eisenstein}, D.~J., {Zehavi}, I., {Hogg}, D.~W., {et~al.} 2005, \apj, 633, 560

\bibitem[{{Eisenstein} {et~al.}(2011){Eisenstein}, {Weinberg}, {Agol},
  {Aihara}, {Allende Prieto}, {Anderson}, {Arns}, {Aubourg}, {Bailey},
  {Balbinot}, \& et~al.}]{eisenstein_sdss3_2011}
{Eisenstein}, D.~J., {Weinberg}, D.~H., {Agol}, E., {et~al.} 2011, \aj, 142, 72

\bibitem[{{Fukugita} {et~al.}(1996){Fukugita}, {Ichikawa}, {Gunn}, {Doi},
  {Shimasaku}, \& {Schneider}}]{fukugita_sdss_1996}
{Fukugita}, M., {Ichikawa}, T., {Gunn}, J.~E., {et~al.} 1996, \aj, 111, 1748

\bibitem[{{Giannantonio} {et~al.}(2014){Giannantonio}, {Ross}, {Percival},
  {Crittenden}, {Bacher}, {Kilbinger}, {Nichol}, \& {Weller}}]{Gian14}
{Giannantonio}, T., {Ross}, A.~J., {Percival}, W.~J., {et~al.} 2014, \prd, 89,
  023511

\bibitem[{{Gunn} {et~al.}(1998){Gunn}, {Carr}, {Rockosi}, {Sekiguchi}, {Berry},
  {Elms}, {de Haas}, {Ivezi{\'c}}, {Knapp}, {Lupton}, {Pauls}, {Simcoe},
  {Hirsch}, {Sanford}, {Wang}, {York}, {Harris}, {Annis}, {Bartozek},
  {Boroski}, {Bakken}, {Haldeman}, {Kent}, {Holm}, {Holmgren}, {Petravick},
  {Prosapio}, {Rechenmacher}, {Doi}, {Fukugita}, {Shimasaku}, {Okada}, {Hull},
  {Siegmund}, {Mannery}, {Blouke}, {Heidtman}, {Schneider}, {Lucinio}, \&
  {Brinkman}}]{gunn_sdss_1998}
{Gunn}, J.~E., {Carr}, M., {Rockosi}, C., {et~al.} 1998, \aj, 116, 3040

\bibitem[{{Gunn} {et~al.}(2006){Gunn}, {Siegmund}, {Mannery}, {Owen}, {Hull},
  {Leger}, {Carey}, {Knapp}, {York}, {Boroski}, {Kent}, {Lupton}, {Rockosi},
  {Evans}, {Waddell}, {Anderson}, {Annis}, {Barentine}, {Bartoszek}, {Bastian},
  {Bracker}, {Brewington}, {Briegel}, {Brinkmann}, {Brown}, {Carr},
  {Czarapata}, {Drennan}, {Dombeck}, {Federwitz}, {Gillespie}, {Gonzales},
  {Hansen}, {Harvanek}, {Hayes}, {Jordan}, {Kinney}, {Klaene}, {Kleinman},
  {Kron}, {Kresinski}, {Lee}, {Limmongkol}, {Lindenmeyer}, {Long}, {Loomis},
  {McGehee}, {Mantsch}, {Neilsen}, {Neswold}, {Newman}, {Nitta}, {Peoples},
  {Pier}, {Prieto}, {Prosapio}, {Rivetta}, {Schneider}, {Snedden}, \&
  {Wang}}]{gunn_sdss_2006}
{Gunn}, J.~E., {Siegmund}, W.~A., {Mannery}, E.~J., {et~al.} 2006, \aj, 131,
  2332

\bibitem[{{Gwyn}(2011)}]{gwyn_2011}
{Gwyn}, S.~D.~J. 2011, ArXiv e-prints, arXiv:1101.1084

\bibitem[{{Ho} {et~al.}(2012){Ho}, {Cuesta}, {Seo}, {de Putter}, {Ross},
  {White}, {Padmanabhan}, {Saito}, {Schlegel}, {Schlafly}, {Seljak},
  {Hern{\'a}ndez-Monteagudo}, {S{\'a}nchez}, {Percival}, {Blanton}, {Skibba},
  {Schneider}, {Reid}, {Mena}, {Viel}, {Eisenstein}, {Prada}, {Weaver},
  {Bahcall}, {Bizyaev}, {Brewinton}, {Brinkman}, {Nicolaci da Costa}, {Gott},
  {Malanushenko}, {Malanushenko}, {Nichol}, {Oravetz}, {Pan},
  {Palanque-Delabrouille}, {Ross}, {Simmons}, {de Simoni}, {Snedden}, \&
  {Yeche}}]{Ho12}
{Ho}, S., {Cuesta}, A., {Seo}, H.-J., {et~al.} 2012, \apj, 761, 14

\bibitem[{{Ilbert} {et~al.}(2008){Ilbert}, {Salvato}, {Capak}, {Le Floc'h},
  {Aussel}, {McCracken}, {Arnouts}, {Mobasher}, {Sanders}, {Scoville}, \&
  {Taniguchi}}]{ilbert_2008}
{Ilbert}, O., {Salvato}, M., {Capak}, P., {et~al.} 2008, in Astronomical
  Society of the Pacific Conference Series, Vol. 399, Panoramic Views of Galaxy
  Formation and Evolution, ed. T.~{Kodama}, T.~{Yamada}, \& K.~{Aoki}, 169

\bibitem[{{Jarrett} {et~al.}(2011){Jarrett}, {Cohen}, {Masci}, {Wright},
  {Stern}, {Benford}, {Blain}, {Carey}, {Cutri}, {Eisenhardt}, {Lonsdale},
  {Mainzer}, {Marsh}, {Padgett}, {Petty}, {Ressler}, {Skrutskie}, {Stanford},
  {Surace}, {Tsai}, {Wheelock}, \& {Yan}}]{wise_sup_jarret_2011}
{Jarrett}, T.~H., {Cohen}, M., {Masci}, F., {et~al.} 2011, \apj, 735, 112

\bibitem[{{John}(1988)}]{john1988}
{John}, T.~L. 1988, \aap, 193, 189

\bibitem[{{Jones}(1968)}]{pogson_mags_1968}
{Jones}, D. 1968, Leaflet of the Astronomical Society of the Pacific, 10, 145

\bibitem[{{Kaiser} {et~al.}(2010){Kaiser}, {Burgett}, {Chambers}, {Denneau},
  {Heasley}, {Jedicke}, {Magnier}, {Morgan}, {Onaka}, \& {Tonry}}]{kaiser_2010}
{Kaiser}, N., {Burgett}, W., {Chambers}, K., {et~al.} 2010, in Society of
  Photo-Optical Instrumentation Engineers (SPIE) Conference Series, Vol. 7733,
  Society of Photo-Optical Instrumentation Engineers (SPIE) Conference Series,
  0

\bibitem[{{Kauffmann} {et~al.}(2004){Kauffmann}, {White}, {Heckman},
  {M{\'e}nard}, {Brinchmann}, {Charlot}, {Tremonti}, \&
  {Brinkmann}}]{Kauffmann2004-353}
{Kauffmann}, G., {White}, S.~D.~M., {Heckman}, T.~M., {et~al.} 2004, \mnras,
  353, 713

\bibitem[{{Lang} {et~al.}(2014){Lang}, {Hogg}, \& {Schlegel}}]{Lan14}
{Lang}, D., {Hogg}, D.~W., \& {Schlegel}, D.~J. 2014, ArXiv e-prints,
  arXiv:1410.7397

\bibitem[{{Leistedt} {et~al.}(2013){Leistedt}, {Peiris}, {Mortlock},
  {Benoit-L{\'e}vy}, \& {Pontzen}}]{Leistedt13}
{Leistedt}, B., {Peiris}, H.~V., {Mortlock}, D.~J., {Benoit-L{\'e}vy}, A., \&
  {Pontzen}, A. 2013, \mnras, 435, 1857

\bibitem[{{Lin} \& {Mohr}(2003)}]{Lin03}
{Lin}, Y.-T., \& {Mohr}, J.~J. 2003, \apj, 582, 574

\bibitem[{{Lupton} {et~al.}(1999){Lupton}, {Gunn}, \& {Szalay}}]{lup_sdss_1999}
{Lupton}, R.~H., {Gunn}, J.~E., \& {Szalay}, A.~S. 1999, \aj, 118, 1406

\bibitem[{{Myers} {et~al.}(2015){Myers}, {et al.}, \& {\em in
  preparation}}]{eboss_qso}
{Myers}, A., {et al.}, \& {\em in preparation}. 2015, to be submitted to \apj

\bibitem[{{Oke} \& {Gunn}(1983)}]{oke_gunn_1983}
{Oke}, J.~B., \& {Gunn}, J.~E. 1983, \apj, 266, 713

\bibitem[{{Padmanabhan} {et~al.}(2008){Padmanabhan}, {Schlegel}, {Finkbeiner},
  {Barentine}, {Blanton}, {Brewington}, {Gunn}, {Harvanek}, {Hogg},
  {Ivezi{\'c}}, {Johnston}, {Kent}, {Kleinman}, {Knapp}, {Krzesinski}, {Long},
  {Neilsen}, {Nitta}, {Loomis}, {Lupton}, {Roweis}, {Snedden}, {Strauss}, \&
  {Tucker}}]{pad_sdss_2008}
{Padmanabhan}, N., {Schlegel}, D.~J., {Finkbeiner}, D.~P., {et~al.} 2008, \apj,
  674, 1217

\bibitem[{{Planck Collaboration} {et~al.}(2014){Planck Collaboration}, {Ade},
  {Aghanim}, {Armitage-Caplan}, {Arnaud}, {Ashdown}, {Atrio-Barandela},
  {Aumont}, {Baccigalupi}, {Banday}, \& et~al.}]{Planck14}
{Planck Collaboration}, {Ade}, P.~A.~R., {Aghanim}, N., {et~al.} 2014, \aap,
  571, A16

\bibitem[{{Postman} \& {Geller}(1984)}]{postman_geller1984}
{Postman}, M., \& {Geller}, M.~J. 1984, \apj, 281, 95

\bibitem[{{Postman} \& {Lauer}(1995)}]{postman_and_lauer1995}
{Postman}, M., \& {Lauer}, T.~R. 1995, \apj, 440, 28

\bibitem[{{Prakash} {et~al.}(2015){Prakash}, {Licquia}, {Newman}, \&
  {Rao}}]{abhi_2015a}
{Prakash}, A., {Licquia}, T.~C., {Newman}, J.~A., \& {Rao}, S.~M. 2015, \apj,
  803, 105

\bibitem[{{Ross} {et~al.}(2011){Ross}, {Ho}, {Cuesta}, {Tojeiro}, {Percival},
  {Wake}, {Masters}, {Nichol}, {Myers}, {de Simoni}, {Seo},
  {Hern{\'a}ndez-Monteagudo}, {Crittenden}, {Blanton}, {Brinkmann}, {da Costa},
  {Guo}, {Kazin}, {Maia}, {Maraston}, {Padmanabhan}, {Prada}, {Ramos},
  {Sanchez}, {Schlafly}, {Schlegel}, {Schneider}, {Skibba}, {Thomas}, {Weaver},
  {White}, \& {Zehavi}}]{Ross11}
{Ross}, A.~J., {Ho}, S., {Cuesta}, A.~J., {et~al.} 2011, \mnras, 417, 1350

\bibitem[{{Ross} {et~al.}(2012){Ross}, {Percival}, {S{\'a}nchez}, {Samushia},
  {Ho}, {Kazin}, {Manera}, {Reid}, {White}, {Tojeiro}, {McBride}, {Xu}, {Wake},
  {Strauss}, {Montesano}, {Swanson}, {Bailey}, {Bolton}, {Dorta}, {Eisenstein},
  {Guo}, {Hamilton}, {Nichol}, {Padmanabhan}, {Prada}, {Schlegel},
  {Maga{\~n}a}, {Zehavi}, {Blanton}, {Bizyaev}, {Brewington}, {Cuesta},
  {Malanushenko}, {Malanushenko}, {Oravetz}, {Parejko}, {Pan}, {Schneider},
  {Shelden}, {Simmons}, {Snedden}, \& {Zhao}}]{ashley2012}
{Ross}, A.~J., {Percival}, W.~J., {S{\'a}nchez}, A.~G., {et~al.} 2012, \mnras,
  424, 564

\bibitem[{{Ross} {et~al.}(2014){Ross}, {Samushia}, {Burden}, {Percival},
  {Tojeiro}, {Manera}, {Beutler}, {Brinkmann}, {Brownstein}, {Carnero}, {da
  Costa}, {Eisenstein}, {Guo}, {Ho}, {Maia}, {Montesano}, {Muna}, {Nichol},
  {Nuza}, {S{\'a}nchez}, {Schneider}, {Skibba}, {Sobreira}, {Streblyanska},
  {Swanson}, {Thomas}, {Tinker}, {Wake}, {Zehavi}, \& {Zhao}}]{Ross14}
{Ross}, A.~J., {Samushia}, L., {Burden}, A., {et~al.} 2014, \mnras, 437, 1109

\bibitem[{{Ross} {et~al.}(2008){Ross}, {Shanks}, {Cannon}, {Wake}, {Sharp},
  {Croom}, \& {Peacock}}]{nic_lrg_2008}
{Ross}, N.~P., {Shanks}, T., {Cannon}, R.~D., {et~al.} 2008, \mnras, 387, 1323

\bibitem[{{Samushia} {et~al.}(2014){Samushia}, {Reid}, {White}, {Percival},
  {Cuesta}, {Zhao}, {Ross}, {Manera}, {Aubourg}, {Beutler}, {Brinkmann},
  {Brownstein}, {Dawson}, {Eisenstein}, {Ho}, {Honscheid}, {Maraston},
  {Montesano}, {Nichol}, {Roe}, {Ross}, {S{\'a}nchez}, {Schlegel}, {Schneider},
  {Streblyanska}, {Thomas}, {Tinker}, {Wake}, {Weaver}, \&
  {Zehavi}}]{Samushia14}
{Samushia}, L., {Reid}, B.~A., {White}, M., {et~al.} 2014, \mnras, 439, 3504

\bibitem[{{Schlafly} \& {Finkbeiner}(2011)}]{snf_sdss2011}
{Schlafly}, E.~F., \& {Finkbeiner}, D.~P. 2011, \apj, 737, 103

\bibitem[{{Schlafly} {et~al.}(2012){Schlafly}, {Finkbeiner}, {Juri{\'c}},
  {Magnier}, {Burgett}, {Chambers}, {Grav}, {Hodapp}, {Kaiser}, {Kudritzki},
  {Martin}, {Morgan}, {Price}, {Rix}, {Stubbs}, {Tonry}, \&
  {Wainscoat}}]{Sch12}
{Schlafly}, E.~F., {Finkbeiner}, D.~P., {Juri{\'c}}, M., {et~al.} 2012, \apj,
  756, 158

\bibitem[{{Schlegel} {et~al.}(1998){Schlegel}, {Finkbeiner}, \&
  {Davis}}]{Sfd1998}
{Schlegel}, D.~J., {Finkbeiner}, D.~P., \& {Davis}, M. 1998, \apj, 500, 525

\bibitem[{{Scranton} {et~al.}(2002){Scranton}, {Johnston}, {Dodelson},
  {Frieman}, {Connolly}, {Eisenstein}, {Gunn}, {Hui}, {Jain}, {Kent},
  {Loveday}, {Narayanan}, {Nichol}, {O'Connell}, {Scoccimarro}, {Sheth},
  {Stebbins}, {Strauss}, {Szalay}, {Szapudi}, {Tegmark}, {Vogeley}, {Zehavi},
  {Annis}, {Bahcall}, {Brinkman}, {Csabai}, {Hindsley}, {Ivezic}, {Kim},
  {Knapp}, {Lamb}, {Lee}, {Lupton}, {McKay}, {Munn}, {Peoples}, {Pier},
  {Richards}, {Rockosi}, {Schlegel}, {Schneider}, {Stoughton}, {Tucker},
  {Yanny}, \& {York}}]{Scranton02}
{Scranton}, R., {Johnston}, D., {Dodelson}, S., {et~al.} 2002, \apj, 579, 48

\bibitem[{{Seo} \& {Eisenstein}(2003)}]{Seo03}
{Seo}, H.-J., \& {Eisenstein}, D.~J. 2003, \apj, 598, 720

\bibitem[{{Smee} {et~al.}(2013){Smee}, {Gunn}, {Uomoto}, {Roe}, {Schlegel},
  {Rockosi}, {Carr}, {Leger}, {Dawson}, {Olmstead}, {Brinkmann}, {Owen},
  {Barkhouser}, {Honscheid}, {Harding}, {Long}, {Lupton}, {Loomis}, {Anderson},
  {Annis}, {Bernardi}, {Bhardwaj}, {Bizyaev}, {Bolton}, {Brewington}, {Briggs},
  {Burles}, {Burns}, {Castander}, {Connolly}, {Davenport}, {Ebelke}, {Epps},
  {Feldman}, {Friedman}, {Frieman}, {Heckman}, {Hull}, {Knapp}, {Lawrence},
  {Loveday}, {Mannery}, {Malanushenko}, {Malanushenko}, {Merrelli}, {Muna},
  {Newman}, {Nichol}, {Oravetz}, {Pan}, {Pope}, {Ricketts}, {Shelden},
  {Sandford}, {Siegmund}, {Simmons}, {Smith}, {Snedden}, {Schneider},
  {SubbaRao}, {Tremonti}, {Waddell}, \& {York}}]{smee_sdss_2013}
{Smee}, S.~A., {Gunn}, J.~E., {Uomoto}, A., {et~al.} 2013, \aj, 146, 32

\bibitem[{{Stoughton} {et~al.}(2002){Stoughton}, {Lupton}, {Bernardi},
  {Blanton}, {Burles}, {Castander}, {Connolly}, {Eisenstein}, {Frieman},
  {Hennessy}, {Hindsley}, {Ivezi{\'c}}, {Kent}, {Kunszt}, {Lee}, {Meiksin},
  {Munn}, {Newberg}, {Nichol}, {Nicinski}, {Pier}, {Richards}, {Richmond},
  {Schlegel}, {Smith}, {Strauss}, {SubbaRao}, {Szalay}, {Thakar}, {Tucker},
  {Vanden Berk}, {Yanny}, {Adelman}, {Anderson}, {Anderson}, {Annis},
  {Bahcall}, {Bakken}, {Bartelmann}, {Bastian}, {Bauer}, {Berman},
  {B{\"o}hringer}, {Boroski}, {Bracker}, {Briegel}, {Briggs}, {Brinkmann},
  {Brunner}, {Carey}, {Carr}, {Chen}, {Christian}, {Colestock}, {Crocker},
  {Csabai}, {Czarapata}, {Dalcanton}, {Davidsen}, {Davis}, {Dehnen},
  {Dodelson}, {Doi}, {Dombeck}, {Donahue}, {Ellman}, {Elms}, {Evans}, {Eyer},
  {Fan}, {Federwitz}, {Friedman}, {Fukugita}, {Gal}, {Gillespie}, {Glazebrook},
  {Gray}, {Grebel}, {Greenawalt}, {Greene}, {Gunn}, {de Haas}, {Haiman},
  {Haldeman}, {Hall}, {Hamabe}, {Hansen}, {Harris}, {Harris}, {Harvanek},
  {Hawley}, {Hayes}, {Heckman}, {Helmi}, {Henden}, {Hogan}, {Hogg}, {Holmgren},
  {Holtzman}, {Huang}, {Hull}, {Ichikawa}, {Ichikawa}, {Johnston}, {Kauffmann},
  {Kim}, {Kimball}, {Kinney}, {Klaene}, {Kleinman}, {Klypin}, {Knapp},
  {Korienek}, {Krolik}, {Kron}, {Krzesi{\'n}ski}, {Lamb}, {Leger},
  {Limmongkol}, {Lindenmeyer}, {Long}, {Loomis}, {Loveday}, {MacKinnon},
  {Mannery}, {Mantsch}, {Margon}, {McGehee}, {McKay}, {McLean}, {Menou},
  {Merelli}, {Mo}, {Monet}, {Nakamura}, {Narayanan}, {Nash}, {Neilsen},
  {Newman}, {Nitta}, {Odenkirchen}, {Okada}, {Okamura}, {Ostriker}, {Owen},
  {Pauls}, {Peoples}, {Peterson}, {Petravick}, {Pope}, {Pordes}, {Postman},
  {Prosapio}, {Quinn}, {Rechenmacher}, {Rivetta}, {Rix}, {Rockosi}, {Rosner},
  {Ruthmansdorfer}, {Sandford}, {Schneider}, {Scranton}, {Sekiguchi}, {Sergey},
  {Sheth}, {Shimasaku}, {Smee}, {Snedden}, {Stebbins}, {Stubbs}, {Szapudi},
  {Szkody}, {Szokoly}, {Tabachnik}, {Tsvetanov}, {Uomoto}, {Vogeley}, {Voges},
  {Waddell}, {Walterbos}, {Wang}, {Watanabe}, {Weinberg}, {White}, {White},
  {Wilhite}, {Wolfe}, {Yasuda}, {York}, {Zehavi}, \&
  {Zheng}}]{stoughton_sdss_2002}
{Stoughton}, C., {Lupton}, R.~H., {Bernardi}, M., {et~al.} 2002, \aj, 123, 485

\bibitem[{{Wright} {et~al.}(2010){Wright}, {Eisenhardt}, {Mainzer}, {Ressler},
  {Cutri}, {Jarrett}, {Kirkpatrick}, {Padgett}, {McMillan}, {Skrutskie},
  {Stanford}, {Cohen}, {Walker}, {Mather}, {Leisawitz}, {Gautier}, {McLean},
  {Benford}, {Lonsdale}, {Blain}, {Mendez}, {Irace}, {Duval}, {Liu}, {Royer},
  {Heinrichsen}, {Howard}, {Shannon}, {Kendall}, {Walsh}, {Larsen}, {Cardon},
  {Schick}, {Schwalm}, {Abid}, {Fabinsky}, {Naes}, \& {Tsai}}]{wright2010}
{Wright}, E.~L., {Eisenhardt}, P.~R.~M., {Mainzer}, A.~K., {et~al.} 2010, \aj,
  140, 1868

\bibitem[{{York} {et~al.}(2000){York}, {Adelman}, {Anderson}, {Anderson},
  {Annis}, {Bahcall}, {Bakken}, {Barkhouser}, {Bastian}, {Berman}, {Boroski},
  {Bracker}, {Briegel}, {Briggs}, {Brinkmann}, {Brunner}, {Burles}, {Carey},
  {Carr}, {Castander}, {Chen}, {Colestock}, {Connolly}, {Crocker}, {Csabai},
  {Czarapata}, {Davis}, {Doi}, {Dombeck}, {Eisenstein}, {Ellman}, {Elms},
  {Evans}, {Fan}, {Federwitz}, {Fiscelli}, {Friedman}, {Frieman}, {Fukugita},
  {Gillespie}, {Gunn}, {Gurbani}, {de Haas}, {Haldeman}, {Harris}, {Hayes},
  {Heckman}, {Hennessy}, {Hindsley}, {Holm}, {Holmgren}, {Huang}, {Hull},
  {Husby}, {Ichikawa}, {Ichikawa}, {Ivezi{\'c}}, {Kent}, {Kim}, {Kinney},
  {Klaene}, {Kleinman}, {Kleinman}, {Knapp}, {Korienek}, {Kron}, {Kunszt},
  {Lamb}, {Lee}, {Leger}, {Limmongkol}, {Lindenmeyer}, {Long}, {Loomis},
  {Loveday}, {Lucinio}, {Lupton}, {MacKinnon}, {Mannery}, {Mantsch}, {Margon},
  {McGehee}, {McKay}, {Meiksin}, {Merelli}, {Monet}, {Munn}, {Narayanan},
  {Nash}, {Neilsen}, {Neswold}, {Newberg}, {Nichol}, {Nicinski}, {Nonino},
  {Okada}, {Okamura}, {Ostriker}, {Owen}, {Pauls}, {Peoples}, {Peterson},
  {Petravick}, {Pier}, {Pope}, {Pordes}, {Prosapio}, {Rechenmacher}, {Quinn},
  {Richards}, {Richmond}, {Rivetta}, {Rockosi}, {Ruthmansdorfer}, {Sandford},
  {Schlegel}, {Schneider}, {Sekiguchi}, {Sergey}, {Shimasaku}, {Siegmund},
  {Smee}, {Smith}, {Snedden}, {Stone}, {Stoughton}, {Strauss}, {Stubbs},
  {SubbaRao}, {Szalay}, {Szapudi}, {Szokoly}, {Thakar}, {Tremonti}, {Tucker},
  {Uomoto}, {Vanden Berk}, {Vogeley}, {Waddell}, {Wang}, {Watanabe},
  {Weinberg}, {Yanny}, {Yasuda}, \& {SDSS Collaboration}}]{york_sdss_2000}
{York}, D.~G., {Adelman}, J., {Anderson}, Jr., J.~E., {et~al.} 2000, \aj, 120,
  1579

\end{thebibliography}

\appendix

\section{Appendix}\label{sec:appendix_a}
\subsection{LRG\_IDROP} \label{sec:appendix_a_idrop}

Objects with LRG-like colors which are too faint for detection in the \textit{i} band but still have a robust detection in the \textit{z}-band can be targeted via a different color-cut. The \textit{r} band photometry for these objects becomes quite noisy and hence it is not used in selection. Instead, we can use a similar selection in a different optical-infrared color-color space:
\begin{align}
      i_{Model}  > 21.8,\\
      z_{Model} \leq 19.5,\label{eqn:z_mod_faint_lrg}\\       
      i-z > 0.7,\label{eqn:iz_faint_lrg}\\
      i-W1 > 2.143 \times (i-z) - 2.0. \label{eqn:iw1_faint_lrg}
\end{align}
Equations~\ref{eqn:iz_faint_lrg} and ~\ref{eqn:iw1_faint_lrg} represent an analogous color selection to equations~\ref{eqn:ri_bright_lrg} and ~\ref{eqn:rw1_bright_lrg}, but using the \textit{i} and \textit{z} bands instead of \textit{r} and \textit{i}. Equation~\ref{eqn:z_mod_faint_lrg} ensures that the objects are well-detected in the \textit{z}-band despite having a noisy detection (if any) in bluer bands. This selection contributes a few targets, $\sim200$ over the entire footprint, which are expected to be at higher redshifts than the standard \eboss\ LRG sample.\\

\section{Appendix}\label{sec:appendix_B}
\subsection{Results from a large pilot survey, SEQUELS} \label{sec:sequels}

As mentioned in \S ~\ref{sec:cat}, the basic ideas underlying the \eboss\ selection algorithm can be implemented in a variety of optical-infrared color spaces. To determine the optimum selection algorithm between two candidate methods, we selected $\sim$70{,}000 LRGs over an area of $\sim$700 \degsq\ with 120.0\degree $< $ $\alpha$ $ <$ 210.0\degree and 45.0\degree $<$ $\delta$ $<$ 60.0\degree. These LRGs were selected by algorithms utilizing two different optical-IR color spaces, and were used to test our selection efficiency and redshift success. The parameters of the selection algorithms were tuned such that one obtains a target density of $\sim$60 \perdegsq\ from each one. In the following sub-sections, we explain the two selection algorithms with their commonalities and major differences.\\

\subsection{Common cuts for \sequels\ LRG samples}\label{sec:sequels_cuts}
First, we require that the RESOLVE\_STATUS bit corresponding to SURVEY PRIMARY is nonzero in order to remove duplicate objects. We also require the photometric flag have the CALIB\_STATUS bit set for all of the \textit{r}, \textit{i}, and \textit{z} bands used for photometric color determinations. In addition, the following flux limits are applied over the entire sample:\\   
\begin{align}
      z_{Fiber2} \leq 21.7, {\rm and}\label{eqn:seq_fiber2_lim}\\
       i_{Model } \geq 19.9,\label{eqn:seq_i_mod_lim}
\end{align}

\subsubsection{\tt {\textbf{r/i/z/WISE LRG selection}}}\label{sec:sequels_cuts_rizwise}   
In the first selection, we identify LRGs using \textit{r-W1}, \textit{r-i} and \textit{i-z} color. This selection algorithm is very similar to the selection described in \S~\ref{sec:target_alg}, differing only due to changes in flux limits to improve completeness. In addition to the common cuts described above, we apply the following selection criteria: 
\begin{align}
      z_{Model} \leq 19.95\label{eqn:z_mod_rizw_lrg},\\       
      r-i > 0.98, \label{eqn:seq_ri_rizw_lrg}\\
      r-W1 > 2.0 \times (r-i), {\rm and}\label{eqn:seq_rw1_rizw_lrg}\\
      i-z > 0.625, \label{eqn:seq_iz_rizw_lrg}
\end{align}
where all variables have the same meanings as in section \ref{sec:mag_lim}. These equations and their relevance have been explained previously in \S~\ref{sec:target_alg}. 
          
\subsubsection{\tt {\textbf{i/z/WISE LRGs }}}\label{sec:sequels_cuts_izwise}  

The second selection is implemented exclusively in \textit{i-W1} and \textit{i-z} optical-IR color-color space, eliminating any use of the \textit{r} band. This selection algorithm is similar to the one explained in \S ~\ref{sec:appendix_a}, differing primarily in its flux limits, which have been tuned to produce the same target density as the \textit{r/i/z/WISE} selection. In addition to the common cuts, we apply the following selection criteria: 
\begin{align}
      z_{Model} \leq 19.5,\label{eqn:z_mod_izw_lrg}\\       
      i-z > 0.7, {\rm and}\label{eqn:iz_izw_lrg}\\
      i-W1 > 2.143 \times (i-z) - 2.0 \label{eqn:iw1_izw_lrg}
\end{align}
The equations and their relevance are the same as explained previously in LRG\_IDROP (\S~\ref{sec:appendix_a}).

\subsection{Details of the \sequels\ survey}
 \sequels\ was conceived as a precursor of \eboss\, enabling us to test the reliability and efficiency of our selection algorithms while simultaneously producing data that could be combined with the full \eboss\ dataset to constrained cosmology. It provided a sufficiently large dataset to enable robust tests of selection algorithms. It was also critical in testing and demonstrating our ability to meet \eboss\ requirements via these selection algorithms. We applied both of the selection algorithms explained in the section \S ~\ref{sec:sequels_cuts} in parallel over the entire \sdss\ footprint. The final \sequels\ LRG sample consisted of the objects selected by either or both of the selection algorithms explained above. 
\subsubsection{Targeting bits}
In order to identify LRGs selected via different algorithms, we assign them different values of the \texttt{eBOSS\_TARGET0} tag. For LRGs selected in \textit{i/z/WISE} color space, \texttt{eBOSS\_TARGET0} is set bit-wise to 1.\footnote{bit 0, 1, 2 are used to indicate $2^{0}=1,\ 2^{1}=2,$ and $2^{2}=4$, respectively.} For LRGs selected via \textit{r/i/z/WISE} selection, \texttt{eBOSS\_TARGET0} is set bit-wise to 2. LRGs which pass both of the selection criteria have both bits set.

\subsubsection{Overall characteristics of \sequels\ LRGs}
The two classes of LRGs, \ie, \textit{r/i/z/WISE} selected and \textit{i/z/WISE} selected, were analyzed separately. We found that $~\geqsim\, $ 87\% of spectra yielded secure redshift measurements. Redshift measurements are checked via visual inspection of the spectra. The remaining 13\% were found to have small differences between the depths of the lowest chi-squared minima, and hence were judged not to be reliable; this generally occurred due to low signal-to-noise ratio in the spectra.  8\% of the total targets were both classified securely and found to be stars. These two factors (13\% of targets having no definitive redshift measurement and another 8\% being stars) make it impossible to reach the required efficiency at targeting $0.6 < z < 1.0$ LRGs of 80\%. We meet the requirement set on the \eboss\ median redshift using the \textit{r/i/z/WISE} algorithm, but not the \textit{i/z/WISE} algorithm. Among the objects which failed to yield a secure redshift measurement, most were noise-dominated. We tabulate the key results in Table~\ref{table:sequels_eboss_req}.\\
 \begin{deluxetable}{lccll}
 \tablecaption{Summary of \textit{r/i/z/WISE} and \textit{i/z/WISE} in comparison to key \eboss\ requirements.}
\tablehead{Requirement & \textbf{\textit{r/i/z/WISE}} & \textbf{\textit{i/z/WISE}} & Summary  \cr 
}
\startdata
 \# of targets: & 450,000  & 450,000 & Easily achievable \cr
  $>375,000$  & ($\sim$ 60 targets \perdegsq) &($\sim$ 60 targets \perdegsq) & & \cr \\
Median Redshift: &\ 0.716 &\ 0.697 &  \textit{i/z/WISE} failing \cr
 $>$0.71& & & marginally\cr\\
 Fraction at $0.6~\leqsim\,z~\leqsim\,1.0$: & $\sim\ 71\%$& $\sim\ 64\%$ & Both samples \cr
$>80$\%  & & & fail to meet \cr
 & & &  \cr
\enddata
\tablecomments{The \textit{r/i/z/WISE} selection meets the basic median redshift requirement which is necessary to achieve our science goals. However, both algorithms fail to meet the redshift efficiency requirement. \textit{r/i/z/WISE} selects more high-redshift LRGs and hence was chosen as the preferred selection algorithm for \eboss.}
\label{table:sequels_eboss_req}
\end{deluxetable}
\begin{figure}[btp]
\centering
\includegraphics[scale=0.77]{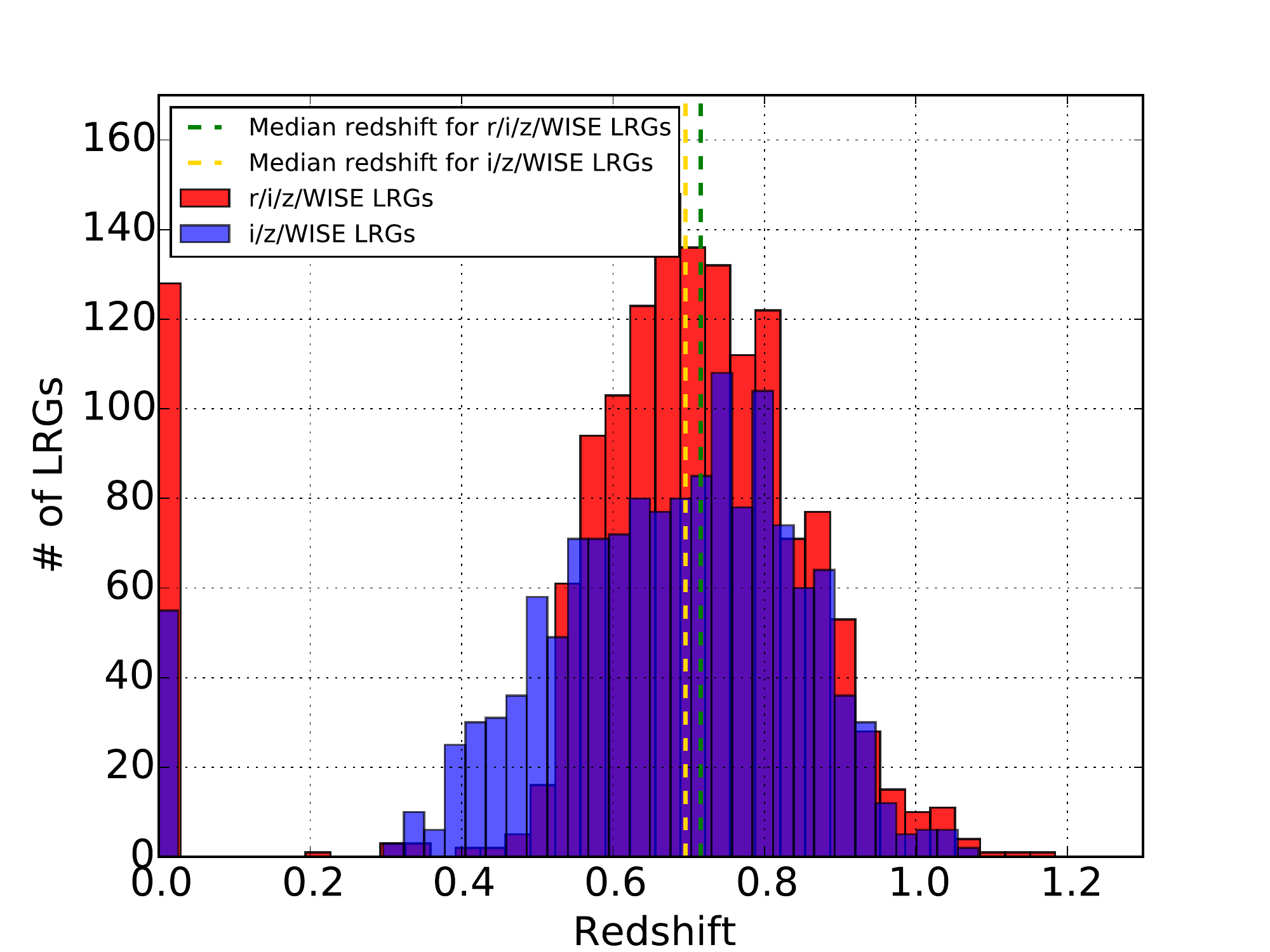}
	\caption{ {Redshift histogram of $\sim$1500 visually inspected LRGs targeted by two different selection algorithms as part of SEQUELS. \textit{r/i/z/WISE} selects more LRGs at higher redshift.  In contrast, \textit{i/z/WISE} selects more LRGs at lower redshifts which are less useful for \eboss. Hence, \textit{r/i/z/WISE} is the preferred choice for the \eboss\ LRG sample.}}
\label{fig:sequels_ri_iz_redshift}
\end{figure}

In Figure~\ref{fig:sequels_ri_iz_redshift} we present the redshift distributions, \textit{N(z)}, of \textit{r/i/z/WISE} and \textit{i/z/WISE} LRGs. We find that the \textit{i/z/WISE} selection algorithm selects a significantly higher fraction of fraction of galaxies at $0.4 < z < 0.5$ compared to the \textit{r/i/z/WISE} selection. This causes the median redshift and targeting efficiency to fall below our requirements, as seen also in Table \ref{table:sequels_eboss_req}. Overall, \textit{r/i/z/WISE}  was found to be more suitable for \eboss.  It gains greater efficiency by requiring targets to be red in both \textit{r-i} and \textit{i-z}, providing a veto in cases where one color is affected by bad photometry.  However, at redshifts $z ~\geqsim\, 0.75$ both of the candidate selection algorithms yielded similar results.

\subsection{Differences between \sequels and \eboss\ targets}
Post SEQUELS, we made a few improvements in our target selection algorithm. These changes are expected to improve our secure redshift measurement rate by removing objects whose counterparts yielded extremely low signal-to-noise spectra in \sequels. For \eboss\ LRGs, we add two additional criteria to the \sequels\ \textit{r/i/z/WISE} selection:

 \begin{align}    
      W1_{AB}  \leq 20.299, {\rm and}\label{eqn:wise_flux_lim_comp}\\
      i_{Model }  \leq 21.8 \label{eqn:i_mod_lim_comp}    
\end{align}
 Equation~\ref{eqn:wise_flux_lim_comp} effectively requires a $5\sigma$ detection in the first channel (\textit{W1}) of \wise. In addition, we put a faint limit on $i_{Modelflux}$ through equation~\ref{eqn:i_mod_lim_comp}; this was not applied in \sequels. These additional flux limits reduce the number of noise-dominated LRG spectra significantly when applied to the \sequels\ sample.

\end{document}